\documentclass[12pt]{article}
\usepackage{amssymb,amsmath,slashed,latexsym}
\usepackage{float}
\allowdisplaybreaks
\newcommand{\ft}[2]{{\textstyle\frac{#1}{#2}}}
\def\Re{\mathop{\rm Re}\nolimits}

\def\rmi{{\rm i}}
\def\rmd{{\rm d}}
\newcommand{\hc}{{\rm h.c.}}
\newcommand{\bbox}{\lower.2ex\hbox{$\Box$}}
\newsavebox{\uuunit}
\sbox{\uuunit}
    {\setlength{\unitlength}{0.825em}
     \begin{picture}(0.6,0.7)
        \thinlines
        \put(0,0){\line(1,0){0.5}}
        \put(0.15,0){\line(0,1){0.7}}
        \put(0.35,0){\line(0,1){0.8}}
       \multiput(0.3,0.8)(-0.04,-0.02){12}{\rule{0.5pt}{0.5pt}}
     \end {picture}}

\csname @addtoreset\endcsname{equation}{section}
\newcommand{\SU}{\mathop{\rm SU}}
\newcommand{\SO}{\mathop{\rm SO}}
\newcommand{\U}{\mathop{\rm {}U}}

   \usepackage[nosort]{cite}
   \pdfoutput=1
  \usepackage[pdftex]{hyperref}
\newcommand{\ib}{\bar{\imath}}
\newcommand{\jb}{\bar{\jmath}}
\newcommand{\poinc}{\boxdot}
\newcommand{\dr}{\raise.3ex\hbox{$\stackrel{\leftarrow}{\delta  }$}{}}
\newcommand{\dl}{\raise.3ex\hbox{$\stackrel{\rightarrow}{\delta }$}{} }
\newcommand{\pl}{\raise.3ex\hbox{$\stackrel{\rightarrow}{\partial }$}{} }

\begin{document}

\begin{titlepage}
\begin{flushright}
CERN-TH-2017-035
\end{flushright}
\vspace{.5cm}
\begin{center}
\baselineskip=16pt
{\LARGE  The Supercurrent and Einstein equations\\ \vskip 0.2cm in the Superconformal formulation  
}\\
\vfill
{\large  {\bf Sergio Ferrara}$^{1,2,3}$, {\bf Marine Samsonyan}$^1$, \\[2mm] {\bf Magnus Tournoy}$^4$ and {\bf Antoine Van Proeyen$^4$}, } \\
\vfill

{\small$^1$ Theoretical Physics Department, CERN CH-1211 Geneva 23, Switzerland\\\smallskip
$^2$ INFN - Laboratori Nazionali di Frascati Via Enrico Fermi 40, I-00044 Frascati, Italy\\\smallskip
$^3$ Department of Physics and Astronomy and Mani L.Bhaumik Institute for Theoretical Physics, U.C.L.A., Los Angeles CA 90095-1547, USA\\\smallskip
$^4$   KU Leuven, Institute for Theoretical Physics, Celestijnenlaan 200D, B-3001 Leuven,
Belgium  \\[2mm] }
\end{center}
\vfill
\begin{center}
{\bf Abstract}
\end{center}
{\small
We give a new expression for the supercurrent and its conservation in curved ${\cal N}=1$, $D=4$ superspace using the superconformal approach. The first component of the superfield, whose lowest component is the vector auxiliary field gives the (super)Einstein equations. Its trace and couplings to conformal and non-conformal matter is presented. In a suitable dilatational gauge, the conformal gauge, we obtain an update of the Callan-Coleman-Jackiw improved currents for conformal matter, containing $R$-symmetry corrections for a new traceless covariantly conserved energy--momentum tensor. We observe that in the Poincar\'{e} gauge, where standard Poincar\'{e} supergravity is usually formulated, the currents are not improved and then the higher conformal symmetry of the matter sector is obscured. The curvature multiplets are used to find supersymmetric curved backgrounds and some examples are exhibited in agreement with existing results.
} \vfill
\begin{center}
\bf  Dedicated to the memory of Aurelio Grillo 
\end{center}
\vfill

\hrule width 3.cm
{\footnotesize \noindent e-mails: Sergio.Ferrara@cern.ch, Marine.Samsonyan@cern.ch, magnus.tournoy@kuleuven.be,\\
antoine.vanproeyen@fys.kuleuven.be }
\end{titlepage}

\addtocounter{page}{1}
 \tableofcontents{}
\newpage
\section{Introduction}

In the superconformal formalism of supergravity, which is a very practical and economic way for an off-shell formulation, the Planck mass $m_{\rm p} = \kappa ^{-1}=  2.4 \times 10^{18}\ \mathrm{GeV}$ emerges as a consequence of the superconformal gauge fixing of a chiral multiplet compensator $X^0$. Alternatively, we can say that a non vanishing compensator spontaneously breaks superconformal to Poincar\'{e}  local supersymmetry.
 A particular interesting application is the case when superconformal matter is present. In this case Callan, Coleman and Jackiw (CCJ)  \cite{Callan:1970ze} showed that a traceless energy-momentum tensor can be defined, which is different for spin 0 and $\ft12$ from the canonical one in that non-minimal couplings of gravity to matter are present. In particular in Poincar\'{e} supergravity the stress tensor is not traceless and the spinor supercurrent is not $\gamma $-traceless. This is only possible if we define a supersymmetric generalization of the improved stress tensor of CCJ. The supercurrent obeying this property satisfies a very simple superfield conservation law \cite{Ferrara:1975pz}\footnote{We use $\approx $ for identities valid due to field equations.}
\begin{equation}
  \bar D^{\dot \alpha }J_{\alpha \dot \alpha }\approx D_\alpha Y\,,\qquad \bar D_{\dot \alpha }Y=0\,,
 \label{conservationJadotaX}
\end{equation}
where $Y\approx 0$ for conformal matter. The vanishing of $Y$ sets to zero  the trace of the energy-momentum tensor, as well as the $\gamma$-trace of the supercurrent and the divergence of the axial current.

At the linearized level it was shown that the supergravity equations in presence of matter correspond to the superfield equation  \cite{Ferrara:1977mv}
\begin{equation}
  \kappa ^{-2} E_{\alpha\dot{\alpha}} + J_{\alpha\dot{\alpha}}\approx 0\,,
 \label{EisJ}
\end{equation}
where $E_{\alpha\dot{\alpha}}$ is the linearized Einstein tensor. Since the latter satisfies a similar conservation equation \cite{Ferrara:1977mv}, which includes the chiral scalar curvature ${\cal R}$,
\begin{equation}
  \bar{D}{}^{\dot \alpha }E_{\alpha \dot \alpha }= D_\alpha {\cal R}\,,
 \label{DE=DR}
\end{equation}
it then follows that
\begin{equation}
  {\cal R}+ \kappa ^2 Y \approx 0\,.
 \label{RS0weak}
\end{equation}
and then ${\cal R}\approx 0$ for conformal matter.
It is the goal of the present investigation to provide fully non-linear expressions for the former quantities using the superconformal techniques \cite{Kaku:1977pa,Ferrara:1977ij,Kaku:1978ea,Kaku:1978nz}, which are explained in \cite{Freedman:2012zz}.
In particular, nonlinearities come from two sources. One is the $X^0$ compensator dependence, and the other is the coupling to matter.
As a by-product we will be able to compute the components of the Einstein multiplet, which will be promoted to a conformal primary superfield together with the scalar curvature.

We will construct the full non-linear Einstein tensor from field equations of pure supergravity. The latter is constructed as the $D$-action of the compensating multiplet $X^0$ (see notations in Appendix \ref{app:superspace}). The field equation for the $R$-symmetry gauge field $A_\mu $ gives the first component of the Einstein tensor multiplet
\begin{equation}
 e^{-1} \frac{\delta }{\delta A^\mu }[-3X^0\bar X^{\bar 0}]_D =-\ft34{\cal E}_\mu  \,,
 \label{calEdefined}
\end{equation}
where ${\cal E}_{\alpha \dot \alpha }$ and ${\cal E}_\mu $ are related as in (\ref{alphaalphadot}).
This is a real superconformal primary field of Weyl weight~3.
The entire supergravity geometry is encoded in the Einstein tensor ${\cal E}_a $, the chiral scalar curvature ${\cal R}$ and the Weyl superfield $W_{\alpha \beta \gamma }$. These objects enter in the superspace formulation of Poincar\'{e} supergravity developed in \cite{Wess:1977fn,Wess:1978bu,Wess:1978ns,Grimm:1978ch}.

The scalar curvature multiplet ${\cal R}$, with chiral and Weyl weights $(1,1)$, is defined in terms of the compensating multiplet as \cite{Kugo:1983mv}
\begin{equation}
  {\cal R}\equiv  \frac{1}{X^0}T(\bar X^{\bar 0})\,,
 \label{Rdefined}
\end{equation}
where the operation $T$ produces a chiral multiplet, and is the local superconformal version of the rigid supersymmetry operation $\bar D^2$.
We will find that the non-linear version of  (\ref{DE=DR}) is that the tensor ${\cal E}_{\alpha\dot{\alpha}}$ and ${\cal R}$ satisfy
\begin{equation}
  \overline{{\cal D}}{}^{\dot \alpha }{\cal E}_{\alpha \dot \alpha }= (X^0)^3 {\cal D}_\alpha \left(\frac{{\cal R}}{X^0}\right)\,,
 \label{sconfeqnER}
\end{equation}
where ${\cal D}_\alpha $ is the superconformal version of the superspace covariant derivative $D_\alpha $
This is the nonlinear generalization of  (\ref{DE=DR}). The latter is a generalization of the identity in general relativity on the Einstein tensor
\begin{equation}
  \nabla ^\mu G_{\mu \nu }=0\,,\qquad G^\mu {}_\mu =-R\,.
 \label{GRanalogue}
\end{equation}
In the absence of matter the Einstein equations for pure supergravity (only the graviton and the gravitino), become
\begin{equation}
 {\cal E}_{\alpha\dot{\alpha}}\approx 0\,,\qquad {\cal R}\approx 0\,,
 \label{pureSUGRAER0}
\end{equation}

In the sequel we will argue that a conformal gauge fixing where\footnote{We indicate equations after gauge fixing to Poincar\'{e} supergravity by the indication $\poinc$.}
\begin{equation}
  \left.X^0\right|_{\poinc} =\kappa ^{-1}\,,
 \label{X0fixed}
\end{equation}
allows us to split gravity and matter in the way of CCJ. Using this gauge fixing, the lowest components of (\ref{pureSUGRAER0}) just reduce to the values of the old-minimal set of auxiliary fields \cite{Ferrara:1978em,Stelle:1978ye,Fradkin:1978jq}
\begin{equation}
  A_a \approx u \approx  0\,,
 \label{ASP0}
\end{equation}
where
\begin{equation}
 \left.{\cal R}\right|_{\poinc} = u\equiv  \kappa \bar F^{\bar 0}\,.
 \label{ERPoinc}
\end{equation}
The equations in (\ref{pureSUGRAER0}) are the superfield generalizations of the pure Einstein equations $G_{\mu \nu}\approx 0$, $R\approx 0$.

When matter is included, the currents are defined similar to (\ref{calEdefined}), using the field equation of the full action for the $R$-symmetry gauge field $A_\mu $.
Prior to the gauge fixing the nonlinear expression on the left-hand side of (\ref{EisJ}) is
\begin{equation}
 e^{-1} \frac{\delta }{\delta A^\mu }{\cal S} ={\cal C}_\mu = -\ft34\left({\cal E}_\mu + J_\mu\right) \,,
 \label{calCdefined}
\end{equation}
where ${\cal S}$ is the action and ${\cal C}_\mu $ and $J_\mu $ are as ${\cal E}_\mu $ real superconformal primary fields of Weyl weight~3.

We will define the concept of the `conformal case' where the action contains a sum of pure supergravity and matter couplings that preserve local conformal invariance. In that case we will find
\begin{equation}
  \overline{{\cal D}}^{\dot \alpha} {\cal E}_{\alpha \dot \alpha}\approx 0,\qquad ({\cal E}_{\alpha \dot \alpha}\,\slashed{\approx }\, 0)\,,\qquad {\cal R}\approx 0\,.
 \label{DERzero}
\end{equation}

The paper is organized as follows. In section \ref{ActionsFieldEquations} we present the superconformal setting and derive the nonlinear form of (\ref{RS0weak}) both for conformal and non-conformal matter.
In section \ref{EinsteinTensorMultiplet} the Einstein tensor multiplet is derived and proven to satisfy (\ref{sconfeqnER}).
In section \ref{SupercurrentMultiplet} we give the supercurrent multiplet for general case, whether the matter is conformal or not. Then we specify two superconformal gauges. One of these corresponds to Einstein frame, and the other to the conformal frame in the sense of  \cite{Callan:1970ze}. It is only in the latter that conformal invariance of the matter system relies in the tracelessness of the energy-momentum tensor and $\gamma $-tracelessness of the supercurrent.
In section \ref{ConformalTensor} we define $W_{\alpha\beta\gamma}$ using the full superconformal curvature $R_{\mu\nu}(Q)$. For on-shell pure supergravity this is the only tensor that specifies the spacetime geometry.
In section \ref{ComponentsofJ} we give the components of the superfields that appear in  (\ref{sconfeqnER}) and (\ref{calCdefined}). Among these components is the Einstein equation in presence of matter and a cosmological constant. In particular, the curvature multiplets give an alternative way to study supersymmetric curved
backgrounds not looking to the gravitino variation but rather to the transformation
properties of these multiplets. In this way we show how the $AdS_4$ and $S^3\times L$
solutions are obtained, in agreement with  \cite{Festuccia:2011ws,Dumitrescu:2012ha}. The consequences of a negative and positive cosmological constant, which follow from (\ref{sconfeqnER}), are also discussed. The bosonic part of our results provide a modification of the improved currents of CCJ, and we provide the full formulae in Sec. \ref{ss:CCJsugra}.
Finally, section \ref{ss:summary} gives some concluding remarks.

In Appendix \ref{FromComptoSuperspace} we give our conventions for superspace quantities and in Appendix \ref{sec:constraints} we recall some aspects of the superconformal Weyl multiplet.
Appendix \ref{app:bosonic} separately discusses the bosonic part of our results, which lead to new improved currents.
In Appendix \ref{ss:methodfe} we give a method that we used in Sec. \ref{ComponentsofJ} to obtain components of superfields in terms of field equations, using invariance of an unspecified action. We obtain there also convenient forms of the Ward identities for all superconformal symmetries. In Appendix \ref{ss:poincmultiplet} we present the Poincar\'{e} form (in conformal gauge) of the Einstein tensor multiplet.

\section{Actions for chiral multiplets and the `conformal case'}\label{ActionsFieldEquations}

The actions of chiral multiplets in the superconformal setup are symbolically obtained from
\begin{equation}
  {\cal S}=\left[N(X,\bar X)\right]_D + \left[{\cal W}(X)\right]_F\,.
 \label{Lconf}
\end{equation}
Here the $X^I$ are superconformal chiral multiplets with Weyl weight~1, and the index $I$ is taken to run over $0,1,\ldots , n$ where $n$ is the number of physical multiplets. The 0-index is included to indicate that one of these multiplets is `compensating' for the superconformal symmetry, such that the physical action will be super-Poincar\'{e} invariant.
The function $N$ is real of Weyl and conformal weights~(2,0), and ${\cal W}$ is holomorphic of Weyl and conformal weights~(3,3). The notation for the actions is explained in Appendix~\ref{app:superspace}:  (\ref{DFactions}) and  (\ref{DFinttheta}). The fields of the Weyl multiplet $\{e_\mu ^a,\,\psi _\mu, \, b_\mu ,\,A_\mu \}$ appear hidden in this notation. We repeat the main ingredients of this multiplet in Appendix~\ref{sec:constraints}.

Pure (Minkowski) supergravity is (with a suitable normalization) obtained for
\begin{equation}
 \mbox{pure supergravity: } N= -3X^0\bar X^{\bar 0}\,,\qquad \Phi = -3\,.
 \label{pureSUGRA}
\end{equation}

For separating pure supergravity from the matter part, it is useful to reorganize the variables $X^I$ in
\begin{equation}
  S^0 = X^0\,,\qquad S^i= \frac{X^i}{X^0}\,,\qquad i=1,\ldots ,n\,,
 \label{introdS0Si}
\end{equation}
such that only $S^0$ has a nonzero Weyl and conformal weights~(1,1). This is the multiplet whose first (complex) component represents the two real auxiliary scalar fields of the old minimal set of auxiliary fields once the superconformal symmetries are gauge-fixed to obtain the super-Poincar\'{e} theory.

The ingredients of the action formula (\ref{Lconf}) are then written as
\begin{equation}
  N(X,\bar X)=S^0\bar S^{\bar 0} \Phi (S,\bar S)\,,\qquad {\cal W}(X)=(S^0)^3W(S)
 \label{actS}
\end{equation}
where by functions of $S$, we understand functions of $S^i$, and we could take out a factor $S^0 \bar S^{\bar 0}$ from $N$ and $(S^0)^3$ from ${\cal W}$, since $N$ is a real function of Weyl weight 2, and ${\cal W}$ is holomorphic of Weyl weight~3.
In view of (\ref{pureSUGRA}) we define the matter coupling function $\Phi _{\rm M}$ by
\begin{equation}
  \Phi (S,\bar S)= -3 + 3\Phi _{\rm M}(S,\bar S)\,,
 \label{Phisplit}
\end{equation}

We define the concept of `\emph{conformal case}' for couplings in which in the $X$-basis the compensator $X^0$ appear in the remainder of $N$ and ${\cal W}$ i.e.
\begin{equation}
  N= -3X^0\bar X^{\bar 0}+ N_{\rm matter}(X^i,\bar X^{\ib})\,,\qquad {\cal W}= W(X^i)\,.
 \label{conformalNW}
\end{equation}
This can be expressed as $N_0 = -3\bar X^{\bar 0}$ and ${\cal W}_0=0$.
Subindexes of $N$ and ${\cal W}$ refer to derivatives w.r.t. the $X^I$, while those of $\Phi $ and $W$ refer to derivatives w.r.t. $S^i$ (with $S^0$ fixed). E.g.
\begin{equation}
    N_0= \frac{\partial }{\partial X^0}N =  \left(\frac{\partial }{\partial S^0} - \frac{S^i}{S^0} \frac{\partial }{\partial S^i}\right)S^0\bar S^{\bar 0} \Phi (S,\bar S) = \bar S^{\bar 0} \left(\Phi - S^i\Phi _i\right)\,.
 \label{N0}
\end{equation}
We can therefore express the difference from the conformal case by quantities
\begin{align}
  \Delta K&\equiv -\frac{1}{3\bar S^{\bar 0}}\left(N_0+3\bar S^{\bar 0}\right)= \frac{1}{3}\left(S^i\Phi _i-\Phi -3\right)= S^i\Phi_{{\rm M}\,i}-\Phi_{\rm M}\,,\nonumber\\
   \Delta W&\equiv \frac{1}{3(S^0)^2}{\cal W}_0= W-\frac{1}{3}S^iW_i\,.
 \label{DeltaKW}
\end{align}
The conformal case is therefore
\begin{equation}
  \mbox{conformal case: }\qquad \Delta K=\Delta W=0\,.
 \label{conformalcase}
\end{equation}
The `conformal case' thus demands that $\Phi _{\rm M}$ should be homogeneous of rank 1 in both $S^i$ and $\bar S^i$, and $W$ homogeneous of rank 3:
\begin{equation}
S^i\Phi_{{\rm M}\,i}=\bar S^{\ib}\Phi_{{\rm M}\,\ib}=\Phi_{\rm M} \,, \qquad S^i W_i=3 W\,.
\label{restrictPhiM}
\end{equation}
Differentiating the first set with respect to $S$ and $\bar S$ one gets
\begin{equation}
  S^i \Phi_{{\rm M} ij}=0 \,, \qquad S^i \Phi_{{\rm M} i \jb}=\Phi_{\rm M \jb}  \,, \qquad S^k \Phi_{{\rm M} k \jb i}=0 \,, \qquad \bar S^{\bar k} \Phi_{{\rm M} \bar k j \ib}=0\,,
\end{equation}
which imply that $\Phi_{{\rm M} i \jb}$ is homogeneous of degree zero.

The simplest case is the $\mathbb{C}\mathbb{P}^n$ model,
\begin{equation}
\Phi=-3+3S^i \bar S^{\ib}=-3+3(X^0\bar X^{\bar 0})^{-1}\, X^i \bar X^{\ib} \,,
 \label{sigmamod}
\end{equation}
and a cubic superpotential. It corresponds to the conformally coupled scalar of \cite{Callan:1970ze}.

In general, if there is a variable $S^1\neq 0$, we can write
\begin{equation}
\Phi _{\rm M}=(X^0\bar X^{\bar 0})^{-1}  X^1  \bar X^{\bar 1} f\left(\frac{X^i}{X^1}, \frac{\bar X^{\ib}}{\bar X^{\bar 1}}\right)=  S^1  \bar S^{\bar1} f\left(\frac{S^i}{S^1}, \frac{\bar S^{\ib}}{\bar S^{\bar 1}}\right)\,.
\end{equation}

Since this action depends on the multiplets $\{X^I\}$ and the Weyl multiplet, the field equations can be divided in those with respect to the matter multiplets and those with respect to the Weyl multiplet. The former will be considered here, and the one for the compensating multiplet will lead to the value of the scalar curvature multiplet ${\cal R}$. The field equations for the Weyl multiplet will define ${\cal E}_\mu$, which will be considered in the next section.

The field equation with respect to the multiplet $X^I$ is the multiplet starting with
\begin{equation}
 e^{-1}\frac{\delta {\cal S}}{\delta X^I}=  \frac12T\left(\frac{\partial N}{\partial X^I}\right)+ \frac{\partial {\cal W}}{\partial X^I}\approx 0\,.
 \label{feXI}
\end{equation}
The operation $T$ is the superconformal version of the superspace operation $\bar D^2$, see Appendix~\ref{FromComptoSuperspace}.
For the compensating multiplet, $I=0$, this can be written as
\begin{equation}
  0\approx \ft12 T(N_0) + {\cal W}_0 = -\ft32\left[ T(\bar S^{\bar 0})+ T(\bar S^{\bar 0}\Delta K) -2 (S^0)^2\Delta W\right]\,.
 \label{feX0}
\end{equation}
The scalar curvature multiplet is defined in (\ref{Rdefined}) as the chiral multiplet with Weyl weight~1:
\begin{equation}
  {\cal R}\equiv \frac{1}{S^0}T\left(\bar S^{\bar 0}\right)\,.
 \label{defR}
\end{equation}
Therefore, the field equation (\ref{feX0}) can be written as
\begin{equation}
  {\cal R}+\frac{Y}{(S^0)^2}\approx 0 \quad \text{with} \quad Y\equiv - 2(S^0)^3 \Delta W +S^0 T(\bar S^{\bar0} \Delta K)\,.
 \label{Rfieldeq}
\end{equation}
In the conformal gauge $S^0= \kappa^{-1}$, we have $ {\cal R}+\kappa^2 Y\approx 0$.
This says that ${\cal R}\approx 0$ for the conformal case. The equation \eqref{Rfieldeq} is related to the global formulae in
\cite{Clark:1995bg,Komargodski:2009rz, Komargodski:2010rb,Kuzenko:2010am,Kuzenko:2010ni,Korovin:2016tsq}, and is the nonlinear version of (\ref{RS0weak}). Our results are valid for the superspace curved geometry described by a chiral compensator $X^0$. Other geometries may correspond to different set of auxiliary fields  \cite{Komargodski:2010rb,Kuzenko:2010am,Kuzenko:2010ni,Korovin:2016tsq,Gates:1983nr,Festuccia:2011ws,Dumitrescu:2012ha}.

\section{Einstein tensor multiplet} \label{EinsteinTensorMultiplet}
We continue with the superconformal formulation, without any gauge fixing so far, and study the Einstein multiplet in this setting and its Bianchi identity (\ref{sconfeqnER}).
We start from the field equation of the field $A_\mu $, which is the gauge field of the $R$-symmetry in the conformal approach, and is the auxiliary field in the super-Poincar\'{e} action:\footnote{We use the notations as in \cite[Ch.17]{Freedman:2012zz}, where the chiral multiplets have components $\{X^I,\,\Omega ^I,\,F^I\}$, with left-handed chiral spinor $\Omega ^I$. We use only chiral multiplets, such that these formulas are a truncation of (17.19-21) in that book.}
\begin{equation}
  e^{-1}\frac{\delta }{\delta A^\mu }[N(X,\bar X)]_D = \rmi N_{\bar I}{\cal D}_\mu \bar X^{\bar I}-\rmi N_I{\cal D}_\mu X^I+\ft{1}{2}\rmi N_{I\bar J}\overline{\Omega }^I\gamma _\mu \Omega ^{\bar J}\,,
 \label{feAmugeneral}
\end{equation}
where
\begin{equation}
  {\cal D}_\mu X^I = \left( \partial _\mu -b_\mu - \rmi A_\mu\right)  X^I
  -\frac{1}{\sqrt{2}}\bar \psi _\mu \Omega ^I\,.
 \label{DmuXI}
\end{equation}
We observe that this expression is invariant under $S$-supersymmetry using
\begin{equation}
  \delta _S{\cal D}_\mu X^I = -\frac{1}{\sqrt{2}}\bar\eta \gamma  _\mu \Omega ^I \,,\qquad \delta _S\Omega ^I  = \sqrt{2}Z P_L\eta \,.
 \label{Stransfos}
\end{equation}
Therefore, this expression is a superconformal primary, and can be used as first component of a superconformal multiplet.

We will identify the Einstein tensor multiplet as the multiplet starting from (a multiple of) this expression in the case of \emph{pure supergravity} (\ref{calEdefined})

Its explicit expression is thus the real vector\footnote{Although it is implicit due to the fact that $\Omega ^I$ is left-handed, we wrote an explicit $P_L =\ft12(1+\gamma _*)$ for clarity.} \cite[(5.5.47)]{Gates:1983nr}
\begin{equation}
  {\cal E}_\mu = 4\rmi X^0{\cal D}_\mu \bar X^{\bar 0}-4\rmi \bar X^{\bar 0}{\cal D}_\mu X^0+2\rmi \overline{\Omega }{}^0P_L\gamma _\mu \Omega ^{\bar 0}\,.
 \label{Emu1}
\end{equation}
which can be written in  the components $\{X^0,\,\Omega ^0,\, F^0\}$ as
\begin{align}
  {\cal E}_\mu &=-8 A_\mu \bar X^{\bar 0}X^0 - 4\rmi \bar X^{\bar 0} \overset{\leftrightarrow}{\partial _\mu} X^0  +2\rmi\bar{\Omega  }^0P_L\gamma _\mu \Omega ^{\bar 0} +2\rmi{\sqrt{2}}\bar \psi _\mu \left(\bar X^{\bar 0} \Omega ^0  -X^0 \Omega ^{\bar 0}\right) \,,\nonumber\\  &\qquad
  \bar X^{\bar 0}\overset{\leftrightarrow}{\partial _\mu}X^0\equiv \bar X^{\bar 0}({\partial _\mu}X^0) -({\partial _\mu}\bar X^{\bar 0})X^0\,.
 \label{Emuconf}
\end{align}
When we go to flat indices, ${\cal E}_a = e_a^\mu {\cal E}_\mu $, this object has Weyl weight 3 and chiral weight~0. This will be important for the generalization of (\ref{DE=DR}).

But before considering the local generalization, let us check that the flat limit of ${\cal E}_\mu $ satisfies  (\ref{DE=DR}).
We identify the superfield with its first component, and write as such a superfield formula (using the notation in Appendix \ref{app:superspace})
\begin{equation}
\begin{split}
  {\cal E}_{\alpha \dot\alpha  }&= \ft14\rmi(\gamma^\mu ) _{\alpha \dot\alpha  }\left.{\cal E}_\mu \right|_{\rm flat}
   \\
   &= - 4 \rmi \bar S^{\bar 0}\overset{\leftrightarrow}{\partial}_{\alpha \dot\alpha }S^0-2(D_\alpha S^0)(\bar D_{\dot\alpha  }\bar S^{\bar 0})\,.
 \label{Ealdotalflat}
 \end{split}
\end{equation}
One can then check that
\begin{equation}
  \bar D^{\dot \alpha } {\cal E}_{\alpha \dot\alpha  } =  (S^0)^3 D_\alpha ((S^0)^{-2}\bar D^2\bar S^{\bar 0})= (S^0)^3 D_\alpha ((S^0)^{-1}{\cal R})\,,
 \label{rigideqn}
\end{equation}
where for the last expression we use the rigid supersymmetry version of (\ref{defR}): ${\cal R}=  (S^0)^{-1}\bar D^2\bar S^{\bar 0} $.

The result is consistent for a generalization as a superconformal formula as mentioned in (\ref{sconfeqnER}). Indeed, to define $\bar D^{\dot \alpha } {\cal E}_{\alpha \dot\alpha  }$ from a vector real superfield, ${\cal E}_{\alpha \dot\alpha  }$ should have Weyl weight~3
following the rules in \cite{Kugo:1983mv}, summarized in \cite[(B.1)]{Ferrara:2016een}. We mentioned already that this is fulfilled with the expression (\ref{Emuconf}). Second, to define the superconformal analogue of $D_\alpha $ on a scalar multiplet, the latter should have $w+c=0$, where $w$ and $c$ are the Weyl and chiral weights. Chiral multiplets satisfy $w=c$, and thus the argument of ${\cal D}_\alpha $ could not be ${\cal R}$, which has Weyl weight zero, but can be $(S^0)^{-1}{\cal R}$. Finally, to match the Weyl weights of the left and right-hand side, the multiplication with a multiplet of $w=c=3$ is imposed. Hence we find that  (\ref{rigideqn})  is possible. To prove that it is indeed fulfilled, we calculate the supersymmetry transformation of ${\cal E}_\mu $:
\begin{equation}
  \delta _Q {\cal E}_a = -\sqrt{2}\rmi \bar \epsilon P_R\left(\gamma _a \Omega ^0 \bar F^{\bar 0} +\slashed{\cal D} X^0\gamma _a \Omega ^{\bar 0} +2 \Omega ^{\bar 0} {\cal D}_a  X^0 - 2 \bar X^{\bar 0}{\cal D}_a\Omega ^0\right)+\hc\,.
 \label{QEa}
\end{equation}
We denote this as
\begin{align}
 \delta {\cal E}_a =& \bar \epsilon P_L \delta _L {\cal E}_a+ \bar \epsilon P_R \delta _R {\cal E}_a\,,\nonumber\\
 &\delta _R {\cal E}_a= -\sqrt{2}\rmi\left(\gamma _a \Omega ^0 \bar F^{\bar 0} +\slashed{\cal D} X^0\gamma _a \Omega ^{\bar 0} +2 \Omega ^{\bar 0} {\cal D}_a  X^0 - 2 \bar X^{\bar 0}{\cal D}_a\Omega ^0\right)\,,
 \label{deltaRE}
\end{align}
and by definition the components of $\delta _R {\cal E}_a$ are the superconformal covariant $\overline{{\cal D}}_{\dot \alpha }{\cal E}_a$. Then
\begin{equation}
  \overline{{\cal D}}^{\dot \alpha }{\cal E}_{\alpha \dot \alpha }= \ft14\rmi(\gamma ^a)_{\alpha \dot \alpha }\overline{{\cal D}}^{\dot \alpha }{\cal E}_a= -\ft14\rmi\left(\gamma ^a\delta _R{\cal E}_a\right)_\alpha \,.
 \label{defDalEaldotal}
\end{equation}
This leads to
\begin{equation}
  \overline{{\cal D}}^{\dot \alpha }{\cal E}_{\alpha \dot \alpha }= \sqrt{2}P_L\left(-\Omega ^0 \bar F^{\bar 0}+ \ft12 X^0\slashed{\cal D}\Omega ^{\bar 0}\right)_\alpha =(X^0)^3\delta_L\left((X^0)^{-2}\bar F^{\bar 0}\right)_\alpha \,.
 \label{simplifieddelE}
\end{equation}
Since $(X^0)^{-1}\bar F$ is the first component of  (\ref{Rdefined}), this confirms (\ref{sconfeqnER}) for the superconformal case.

Finally, let us consider the super-Poincar\'{e} expressions. For the pure supergravity case, we can use the gauge fixing of the extra symmetries in the superconformal algebra as
\begin{equation}
  \left.X^0\right|_\poinc = \kappa^{-1} \,,\qquad \left.\Omega ^0 \right|_\poinc = 0\,,\qquad \left.b_\mu \right|_\poinc = 0\,.
 \label{PoincZchi}
\end{equation}
This implies
\begin{equation}
  \left.{\cal E}_a\right|_\poinc = -8\kappa ^{-2} A_a\,.
 \label{EaPoinc}
\end{equation}

Thus in this section, we found the Einstein tensor multiplet as the supercurrent multiplet for the compensator $X^0$ as in (\ref{Emu1}), which corresponds to \cite[(32)]{Ferrara:1975pz}.

\section{The supercurrent multiplet} \label{SupercurrentMultiplet}

The advantage of the superconformal tensor calculus is that it puts on equal footing the compensator $X^0$ and the physical matter chiral multiplets in the collection $\{X^I\}$, all with conformal and chiral weight $(1,1)$. Therefore, the $A_\mu $ field equation should read as
\begin{equation}
  e^{-1}\frac{\delta }{\delta A^\mu }[N(X,\bar X)]_D = -\ft34 ({\cal E}_\mu+ J_\mu )\,,
 \label{E+Jdef}
\end{equation}
Since the superpotential contribution $[{\cal W}]_F$ does not involve $A_\mu $, we have by definition
\begin{equation}
  {\cal E}_\mu+ J_\mu \approx 0\,.
 \label{E+Japprox0}
\end{equation}
The expression ${\cal E}_a+ J_a $ is still a conformal primary of Weyl weight~3. Hence the operations of the previous section are well defined in the conformal setting. We obtained  (\ref{sconfeqnER})
and therefore
\begin{equation}
\overline{{\cal D}}^{\dot\alpha} {\cal J}_{\alpha\dot\alpha}\approx (X^0)^3 {\cal D}_\alpha \left(\frac{Y}{(X^0)^3}\right)\,.
\end{equation}

From (\ref{feAmugeneral}) we thus find
\begin{equation}
  -\ft34 ({\cal E}_\mu+ J_\mu )= \rmi N_{\bar I}{\cal D}_\mu \bar X^{\bar I}-\rmi N_I{\cal D}_\mu X^I+\ft{1}{2}\rmi N_{I\bar J}\overline{\Omega }^I\gamma _\mu \Omega ^{\bar J}\,.
 \label{EplusJ}
\end{equation}
Since  this expression is linear in $N$, after the split (\ref{Phisplit}), and
before elimination of auxiliary fields and conformal gauge fixing, we then have
\begin{equation}
  J_\mu = -4 e^{-1}\frac{\delta }{\delta A^\mu }[X^0\bar X^{\bar 0}\Phi_{\rm M} (S,\bar S)]_D\,.
 \label{JmuinPhiM}
\end{equation}

From (\ref{E+Japprox0}) we obviously also have
\begin{equation}
  \overline{ {\cal D}}^{\dot \alpha }\left({\cal E}_{\alpha \dot \alpha }+ J_{\alpha \dot \alpha }\right)\approx 0\,.
 \label{conservationE+J}
\end{equation}
Using the identity (\ref{sconfeqnER}), this also implies, see (\ref{Rfieldeq}),
\begin{equation}
\begin{split}
 \overline{{\cal D}}^{\dot \alpha }J_{\alpha \dot \alpha }&\approx - (S^0)^3 {\cal D}_\alpha ((S^0)^{-1}{\cal R})\\
 &\approx - (S^0)^3 {\cal D}_\alpha \left(2\Delta W -(S^0)^{-2}T\left(\bar S^{\bar0}\Delta K\right)\right)\,.
 \label{conservJ}
 \end{split}
\end{equation}
In the conformal case,
\begin{equation}
  \overline{{\cal D}}^{\dot \alpha }J_{\alpha \dot \alpha }\approx 0\,.
 \label{confconsJ}
\end{equation}

The expression for $J_{\alpha \dot \alpha }$ from (\ref{EplusJ}) can be written in terms of the quantities that we defined in Sec. \ref{ActionsFieldEquations}. We will thus use the variables $S^i$, and translate e.g.
\begin{equation}
  N_0= \frac{\partial }{\partial X^0}N = \left(\frac{\partial}{\partial S^0}-\frac{S^i}{S^0}\frac{\partial }{\partial S^i}\right)N = \bar S^{\bar 0}\left(\Phi -S^i\Phi _i\right)\,.
 \label{N0translated}
\end{equation}
Using further $\Phi _{\rm M}$ and $\Delta K$, defined in  (\ref{Phisplit}) and  (\ref{DeltaKW}) and the real quantity
\begin{equation}
  \bar \Delta \Delta K = \Phi_{\rm M}- S^i \Phi_{{\rm M}\,i} -\bar S^{\ib} \Phi_{{\rm M}\,\ib} + S^i\bar S^{\jb} \Phi_{{\rm M}\,i\jb}= \left(\bar S^{\ib}\frac{\partial }{\partial \bar S^{\ib}}-1\right) \Delta K\,,
 \label{DeltaDeltaK}
\end{equation}
we have
\begin{equation}
  \begin{split}
  &N_0= -3\bar S^{\bar 0}(1+\Delta K)\,,\qquad N_i = 3 \bar S^{\bar 0}\Phi_{{\rm M}\,i}\,,\\
  &N_{0\bar 0}= -3 +3 \bar \Delta \Delta K\,,\qquad N_{0\ib}= -3 \frac{\partial }{\partial \bar S^{\ib}}\Delta K\,,\qquad N_{i\jb}=3\Phi _{{\rm M},i\jb}\,. \label{NtoPhiM}
  \end{split}
\end{equation}
We then use  $\chi ^i=P_L\chi ^i$ for the fermionic partner of $S^i$. This gives
\begin{equation}
\begin{split}
  &{\cal D}_\mu X^i = S^i{\cal D}_\mu X^0 + X^0{\cal D}_\mu S^i\,,\qquad {\cal D}_\mu S^i= \partial _\mu S^i -\frac{1}{\sqrt{2}}\bar \psi _\mu \chi ^i\,,\\
  &\Omega ^0=\chi ^0\,,\qquad \Omega ^i = X^0\chi ^i +S^i\chi ^0\,.
  \end{split}
 \label{calDmuXSi}
\end{equation}
We therefore find
\begin{eqnarray}
  J_\mu  &=&-\Phi _{\rm M}{\cal E}_\mu  \nonumber\\
  &&+2\rmi \bar X^{\bar 0}\Phi _{{\rm M}\,\ib}\bar \chi ^{\bar 0}\gamma _\mu \chi ^{\ib}
  +2\rmi X^0\Phi _{{\rm M}\,i}\bar \chi ^i\gamma _\mu \chi ^{0}\nonumber\\
&&  + 2\rmi X^0\bar X^{\bar 0}\left[2(\Phi _{{\rm M}\,i}{\cal D}_\mu S^i-\Phi _{{\rm M}\,\ib}{\cal D}_\mu \bar S^{\ib})-\Phi _{{\rm M}\,i\jb}\bar\chi ^i\gamma _\mu \chi ^{\jb} \right]\,.
 \label{EJPhi}
\end{eqnarray}

We will compare two different gauge fixings of dilatational and $S$-supersymmetry. The Einstein gauge is obtained by
\begin{align}
 \mbox{Einstein gauge :}\quad & -3\kappa ^{-2} =\left. N \right|_\poinc =\left. 3X^0\bar X^{\bar 0} \left(-1 +  \Phi_{\rm M}\right) \right|_\poinc \,,\nonumber\\
 & 0=\left.N_I\Omega ^I\right|_\poinc = \left.3\bar X^{\bar 0}\left[(-1+\Phi_{\rm M})\chi ^0 +  \Phi_{{\rm M}\,i}\chi ^i\right]\right|_\poinc\,.
 \label{gfN}
\end{align}
This leads to
\begin{equation}
  \left. X^0\bar X^{\bar 0}\right|_\poinc= \kappa ^{-2}\left(1-\Phi_{\rm M} (S,\bar S)\right)^{-1}\,,\qquad \chi ^0 = \Phi_{{\rm M}\,i}\chi ^i\left(1-\Phi_{\rm M} (S,\bar S)\right)^{-1}\,.
 \label{X0fix}
\end{equation}
Therefore the supergravity and matter fields in ${\cal E}_\mu $ in (\ref{Emu1}) get mixed.

Instead, we can consider the following gauge choice
\begin{equation}
 \mbox{Conformal gauge :}\quad\left.X^0\right|_\poinc=\kappa^{-1}\, , \qquad \left.\Omega^0\right|_\poinc=0\, . \label{gfconf}
\end{equation}
In this gauge choice the compensating $S$-transformations to stay in the gauge only involve the multiplet $\{X^0,\Omega ^0,F^0\}$ and the Weyl multiplet background. But the multiplets $\{S^i,\chi ^i, F^i\}$ do not enter in the transformations of $\Omega ^0$ and thus not in the decomposition rule. Hence, the transformations do not mix the gravity with matter.
 In this gauge ${\cal E}_\mu $ does not depend on matter fields, and we have
\begin{align}
\left.  {\cal E}_\mu \right|_\poinc&= -8\kappa ^{-2} A_\mu\,,\nonumber\\
\left.  J_\mu  \right|_\poinc &=8\kappa ^{-2} \Phi_{\rm M} A_\mu +2\rmi\kappa ^{-2}\left[2(\Phi _{{\rm M}\,i}{\cal D}_\mu S^i-\Phi _{{\rm M}\,\ib}{\cal D}_\mu \bar S^{\ib})-\Phi _{{\rm M}\,i\jb}\bar\chi ^i\gamma _\mu \chi ^{\jb} \right]\,.
 \label{EJPoincconfgauge}
\end{align}
We thus find here the generalization of the equations obtained in \cite{Callan:1970ze} for conformal matter.
The matter action contains a term $R \Phi$, such that the coupling is conformal.
The bosonic part of these equations lead to improved currents and a modified Einstein equation with a
matter energy-momentum tensor that contains the gravity part $G_{\mu \nu }\Phi _{\rm M}$ and a $\U(1)$ part, such that it is conserved and traceless due to the equations of motions. This is different from the
Einstein gauge, (\ref{X0fix}), where $\Phi$ decouples from $R_{\mu\nu}$, but the energy-momentum tensor is not traceless. These bosonic results are discussed in Appendix \ref{app:bosonic}.

The references \cite{Townsend:1979js,Ferrara:1988qx} derived the gravitational multiplets directly with the Poincar\'{e} calculus. This corresponds to our conformal gauge and it explains why the transformations are matter independent. Their formulae agree with ours provided the Poincar\'{e} chiral curvature $R$ is identified with our (zero Weyl and chiral weight) ${\cal R}/X^0$ expression in \eqref{calRoverX0}.

\section{The superconformal tensor} \label{ConformalTensor}
The Weyl superconformal tensor is a 3-spinor index chiral quantity $W_{\alpha \beta \gamma }$  \cite{Ferrara:1977mv,Wess:1978ns,Grimm:1978ch}. It is compensator-independent and the multiplet defined by this field contains in its bosonic components the Weyl tensor and the field strength of the $R$-symmetry vector.
This multiplet has been given in the superconformal context in \cite{Townsend:1979js,Ferrara:1988qx}, but for completeness we repeat it here using our present conventions.\footnote{In the framework of new-minimal Poincar\'{e} calculus, this multiplet was constructed in \cite{Ferrara:1988qxa,Ferrara:1988pd}.} We start from the full curvature $R_{\mu \nu }(Q)$, which is (see Appendix \ref{sec:constraints})
\begin{equation}
  R_{\mu \nu }(Q)  =R'_{\mu \nu } -2\gamma_{[\mu }\phi _{\nu ]}\,,
 \label{RQcomplete}
\end{equation}
and satisfies
\begin{equation}
  \gamma ^\mu R_{\mu \nu }(Q)= \gamma ^\mu \tilde R_{\mu \nu }(Q)=0\,.
 \label{RQprop}
\end{equation}
We define
\begin{equation}
 W_{\alpha \beta \gamma }= (\gamma^{\mu \nu })_{\alpha \beta }R_{\mu \nu }(Q)_\gamma \,,
 \label{WRQ}
\end{equation}
where the spinor indices indicate a $P_L$ projection. By the properties of gamma matrices in 4 dimensions, this is symmetric in $(\alpha \beta )$, as we now prove. We perform a Fierz transformation 
\begin{equation}
  W_{\alpha \beta \gamma }=\ft12 {\cal C}_{\gamma \beta }(\gamma ^{\mu \nu }  R_{\mu \nu }(Q))_\alpha
 - \ft18(\gamma ^{\rho \sigma })_{\gamma \beta  }(\gamma ^{\mu \nu }\gamma _{\rho \sigma } R_{\mu \nu }(Q))_\alpha  \,.
 \label{WFierz}
\end{equation}
After using (\ref{RQprop}) and $\gamma $-algebra, this gives
\begin{equation}
   W_{\alpha \beta \gamma }=(\gamma ^{\rho \sigma })_{\gamma \beta }R_{\rho \sigma }(Q)_\alpha  \,.
 \label{WafterFierz}
\end{equation}
Hence this is indeed symmetric in the three spinor indices. Since $R_{\mu \nu }(Q)$ is invariant under $S$-supersymmetry, this is also a superprimary. The $Q$-supersymmetry transformation is
\begin{equation}
\delta_Q P_L R_{ab}(Q)=\left(\ft14 R_{ab}^{\rm cov}(M^{cd})\gamma_{cd}-\rmi  R_{ab}^-(T)\right)P_L\epsilon\,,
\label{varfermcomp}
\end{equation}
where ${R}^{\rm cov}$ is defined in (\ref{depfields}).
Therefore $W_{\alpha \beta \gamma }$ is a superconformal chiral multiplet with (Weyl, chiral) weight $(\ft32,\ft32)$. Its next component is defined from
\begin{equation}
  \delta W_{\alpha\beta\gamma}=\frac{1}{\sqrt 2} \epsilon^\delta C_{\alpha\beta\gamma\delta}\,,
 \label{delWC}
\end{equation}
where all the spinor indices are left-chiral. This leads to
\begin{equation}
C_{\alpha\beta\gamma\delta}=-\sqrt2(\gamma^{ab})_{\alpha\beta}\left(\ft14 R_{ab}^{\rm cov}(M^{cd})(P_L\gamma_{cd})-\rmi R^-_{ab}(T)(P_L)\right)_{\gamma\delta}\,.
\label{Ctensor}
\end{equation}
Using the constraint in (\ref{eq:constraints}) one proves that this tensor is also traceless.

\section{Components of current multiplets}\label{ComponentsofJ}

In this section we will obtain information on the components of the conformal current multiplets ${\cal E}_a$ and $J_a$. They are defined from field equations of the $R$-symmetry gauge field in  (\ref{calCdefined}) and (\ref{calEdefined}). These multiplets are real conformal multiplets with a vector index of Weyl weight~3. The components can be determined from the $Q$-supersymmetry transformations of their lowest components.

\subsection{The real vector multiplet}

The dilatation transformation $D$, the $\U(1)$ transformation $T$,  $Q$ and $S$ supersymmetry and special conformal transformations $K$ for a real vector multiplet with Weyl weight $w$ are (see \cite{Kugo:1983mv} and in Poincar\'{e} tensor calculus \cite{Townsend:1979js,Ferrara:1988qx})
\begin{align}
\delta\mathcal{C}_{a}=&\;w\lambda _{\rm D}\mathcal{C}_{a} +\ft{1}{2}\rmi\bar{\epsilon}\gamma_{*}\mathcal{Z}_{a}\,,\nonumber\\
\delta P_{L}\mathcal{Z}_{a}=&\left[(w+\ft12)\lambda _{\rm D}-\ft32\rmi\lambda _T\right]P_{L}\mathcal{Z}_{a}\nonumber\\
&+ \ft{1}{2}P_{L}\left(\rmi\mathcal{H}_{a}-\gamma^{b}\mathcal{B}_{ba}-\rmi\slashed{\mathcal{D}}\mathcal{C}_{a}\right)\epsilon+\rmi P_L \left(-w{\cal C}_a+\gamma _{ab} {\cal C}^b\right)\eta\,,\nonumber\\
\delta\mathcal{H}_{a}=&\left[(w+1)\lambda _{\rm D}-3\rmi\lambda _T\right]\mathcal{H}_{a}\nonumber\\
&-\rmi\bar{\epsilon}P_{R}\left(\slashed{\mathcal{D}}\mathcal{Z}_{a}+\Lambda_{a}\right)+\rmi\bar \eta P_L\left((w-2){\cal Z}_a+\gamma _{ab}{\cal Z}^b\right)\,,\nonumber\\
\delta\mathcal{B}_{ba}=&\left[(w+1)\lambda _{\rm D}\right]\mathcal{B}_{ba}+2\varepsilon _{abcd}{\cal C}^c\lambda _{\rm K}^d\nonumber\\
&-\ft{1}{2}\bar{\epsilon}\left(\mathcal{D}_{b}\mathcal{Z}_{a}+\gamma_{b}\Lambda_{a}\right)+ \ft12\rmi \bar{R}_{ac} (Q)\gamma_* \gamma_b \epsilon \,\mathcal{C}^c+\ft12\bar \eta \left((1+w)\gamma _b{\cal Z}_a+ \gamma _{ac}\gamma _b{\cal Z}^c\right)\nonumber\\
\delta P_{R}\Lambda_{a}=&\left[(w+\ft32)\lambda _{\rm D}-\ft32\rmi\lambda _T\right]P_{R}\Lambda_{a}+w P_R\slashed{\lambda }_{\rm K}{\cal Z}_a-\gamma _{ab}\slashed{\lambda }_{\rm K}{\cal Z}^b \nonumber\\
&+\ft{1}{2}\left[\gamma^{bc}\left(\mathcal{D}_{b}\mathcal{B}_{ca}+\rmi\mathcal{D}_{b}\mathcal{D}_{c}\mathcal{C}_{a}\right)-\rmi D_{a}\right]P_{R}\epsilon-\ft{1}{2}P_R \gamma^d \epsilon \bar{R}_{ab}(Q)P_R\gamma_d {\cal Z}^b\nonumber\\
&+\ft12P_R\left(\rmi{\cal H}_b-\gamma ^c{\cal B}_{cb}+\rmi \slashed{\cal D}{\cal C}_b\right)\left(w\delta ^b_a-\gamma _a{}^b\right)\eta \,,\nonumber\\
\delta D_{a}=&\left[(w+2)\lambda _{\rm D}\right] D_{a}+2w\lambda_K^b{\cal D}_b{\cal C}_a+4\lambda_{K[a}{\cal D}_{b]}{\cal C}^b-2\varepsilon_{abcd}{\cal B}^{cd}\lambda_K^b \nonumber\\
&+\ft{1}{2}\rmi\bar{\epsilon}\gamma_{*}\slashed{\mathcal{D}}\Lambda_{a}+\bar \epsilon \left(R_{ab}(T)+\gamma_* \tilde{R}_{ab}(T)\right){\cal Z}^b-\ft{1}{2}\bar{\epsilon}\left(\rmi \gamma_* \gamma _d{\cal B}^{dc} -\slashed{\cal D}{\cal C}^c\right){R}_{ac}(Q)\nonumber\\
&+\rmi w\bar \eta \gamma _*\left( \Lambda_a +\ft12\slashed{\cal D}{\cal Z}_a \right)+\rmi\bar \eta \gamma _*\gamma _{ab}\left(\Lambda ^b+\ft12\slashed{\cal D}{\cal Z}^b\right) \,.
\label{vectorreal}
\end{align}
Since ${\cal C}_\mu $ is defined as a field equation, see (\ref{calCdefined}), it transforms to field equations, and thus the next components of the multiplet are also combinations of field equations. In Appendix \ref{ss:methodfe} we explain how we can use general equations for transformations of equations of motion to identify these further components (based on \cite{Vanhecke:prep}). This leads to expressions that do not depend on the particular action, as long as one considers an action that is invariant under all the symmetries of the superconformal group. The result then depends on `covariantized field equations'
\begin{align}
&\Theta (A)_a = e^{-1}e_a^\mu\frac{\delta {\cal S}}{\delta A^{\mu}} \,,\qquad
  \Theta (e)_{ba}= e^{-1}e_b^\nu \frac{\delta {\cal S}}{\delta e^{\nu a}}+\ldots \,,\qquad \Theta (\psi )_a= e^{-1}e_{a}^\mu \frac{\dl {\cal S}}{\delta \bar \psi ^\mu }+\ldots \,,\nonumber\\
&\Theta (F)_I = e^{-1}  \frac{\delta {\cal S}}{\delta F^I}\,,\qquad \Theta (\Omega )_I= e^{-1}\frac{\dl {\cal S}}{\delta\bar \Omega ^I}+\ldots \,,\qquad \Theta (X)_I= e^{-1}  \frac{\delta {\cal S}}{\delta X^I}+\ldots \,,
 \label{Thetadefs}
\end{align}
where the $\ldots $ make the expression covariant. The expressions for $\Theta (A)_a$ and $\Theta (F)_I$ are covariant without extra terms as can be understood from the general principles in (\ref{ThetaifromH}).
The result for the components of such real multiplet is
\begin{align}
\mathcal{C}_{a}=&\;\Theta(A)_{a},\nonumber\\
\mathcal{Z}_{a}=&\;  3\Theta(\psi)_{a}-\gamma _a\gamma ^b\Theta(\psi)_{b} +\frac{1}{\sqrt{2}}\gamma _a\left(\Omega^{I}\Theta(F)_{I}+\hc\right)\,,\nonumber\\
\mathcal{H}_{a}=&\;2\rmi \bar{X}^{\bar{I}}\mathcal{D}_{a}\Theta(F)_{\bar{I}}-4\rmi \mathcal{D}_{a}\bar{X}^{\bar{I}}\Theta(F)_{\bar{I}},\nonumber\\
\mathcal{B}_{ba}=&\;3\Theta(e)_{ab}-\eta_{ab}\Theta(e)_c{}^c+\ft{1}{2}\varepsilon_{abcd}\mathcal{D}^{c}\Theta(A)^{d}-\eta_{ab}\left(\ft12\bar{\Omega}^{I}\Theta(\Omega)_{I}+F^{I}\Theta(F)_{I}+\hc\right)\,,\nonumber\\
\Lambda_{a}=&\;2\gamma^b {\cal D}_{[a} {\cal Z}_{b]}-3\sqrt{2}\left(\Theta(\Omega)^I {\cal D}_a X_I +\Omega^I {\cal D}_a \Theta(F)_I + \hc\right)\,,\nonumber\\
D_{a}=&-2{\cal D}^b{\cal D}_{[b}\Theta(A)_{a]}-2{\cal D}_{[a}{\cal D}_{b]}\Theta(A)^b\nonumber\\
&-\ft32 \rmi\left(2\mathcal{D}_{a}X^{I}\Theta(X)_{I}-\bar{\Omega}^{I}\overset{\leftrightarrow}{\mathcal{D}}_{a}\Theta(\Omega)_{I}-2F^{I}\mathcal{D}_{a}\Theta(F)_{I}-\hc\right)\,. \label{ETM}
\end{align}

We can apply these results first for the \emph{pure supergravity} action $[-3X^0\bar X^{\bar 0}]_D$, and as defined in (\ref{calEdefined}) this leads to the components of the superfield ${\cal E}$. In this case the covariant field equations of the compensating multiplet are simple:
\begin{equation}
  \Theta(X)_0=-3\bbox^c \bar{X}^{\bar 0}\,,\qquad \Theta(\Omega)_0=3P_L\slashed{\cal D}\Omega^{\bar 0}\,,\qquad \Theta(F)_0=-3\bar{F}^{\bar 0}\,.
 \label{Thetacompensating}
\end{equation}
These $\{\Theta(F)_0,\, -\Theta(\Omega)_0,\,\Theta(X)_0\}$ form a chiral multiplet of Weyl weight~2, which is $-3 X^0\,{\cal R}$.
The covariant field equations for the fields of the Weyl multiplet are (see also (\ref{Emu1})):
\begin{align}
  \Theta (A)_a =&  - 3\rmi X^0{\cal D}_a \bar X^{\bar 0}+3\rmi \bar X^{\bar 0}{\cal D}_a X^0-\ft32\rmi \overline{\Omega }{}^0P_L\gamma _a \Omega ^{\bar 0}\,,\nonumber\\
  \sqrt{2}  \Theta (\psi )_a =& -3P_L\Omega^0 ({\cal D}_a \bar X^{\bar 0})+P_L\gamma _{ab}\left(\Omega^0 {\cal D}^b \bar X^{\bar 0} -2 \bar X^{\bar 0} {\cal D}^b\Omega^0 \right)+\hc
  \nonumber\\
  \Theta (e)_{ab} =&\;3\eta _{ab}\left[F^0\bar F^{\bar 0} - {\cal D}_cX^0{\cal D}^c\bar X^{\bar 0} -\ft12\bar \Omega^0  P_L\slashed{\cal D}\Omega^{\bar 0} -\ft12\bar \Omega^{\bar 0}
 P_R\slashed{\cal D}\Omega^0\right] \nonumber\\
   &   +6 {\cal D}_{(a} X^0\,{\cal D}_{b)}\bar X^0 +\left(\eta _{ab}{\cal D}^c{\cal D}_c-{\cal D}_{(a} {\cal D}_{b)}\right) (X^0\bar X^0)\nonumber\\
& +\ft32\left[\bar \Omega^0 P_L\left(\gamma _{(a}{\cal D}_{b)} -\ft12\gamma _{ab} \slashed{\cal D}\right)\Omega^{\bar 0} +\hc\right] \,.
\label{ThetaWeyl}
\end{align}

When we go to conformal gauge (\ref{gfconf}), the remaining objects combine in covariant Poincar\'{e} covariant quantities.
With the Poincar\'{e} gauges (\ref{PoincZchi}) the Poincar\'{e} supersymmetry transformations are (and it is now convenient to use again $u= \kappa \bar F^0$ as in (\ref{ERPoinc}))
\begin{align}
  \delta _{\poinc}(\epsilon )=& \delta _Q(\epsilon )+ \delta _S\left(\eta = \ft12(\rmi\gamma _*\slashed{A}-P_R u-P_L\bar u)\epsilon \right)+\delta _{\rm K}\left(\lambda _{{\rm K}a}= -\ft14\bar \epsilon \hat{\phi }_a \right)\,,\nonumber\\
  &P_L\hat{\phi }_a=P_L\phi _a+\ft12 P_L(\rmi\slashed{A}+\bar u)\psi _\mu \,.
 \label{delPoincare}
\end{align}
The covariant derivatives
\begin{equation}
  \widehat{{\cal D}}_\mu = \partial _\mu- \delta _M\left(\omega _\mu {}^{ab}(e,\psi )\right) -\delta _{\poinc}(\psi _\mu  )\,,
 \label{PoinccalD}
\end{equation}
are now covariant for super-Poincar\'{e}. Relevant Poincar\'{e} curvatures\footnote{Note that due to the torsion, $R_{ab}$ is not symmetric. However, the covariantized $\widehat{R}_{ab}$ is symmetric.} and derivatives are
\begin{align}
  P_L\widehat{R}_{\mu \nu  }(Q) & = 2P_L\left( \partial _{[\mu }-\ft32\rmi
A_{[\mu }\gamma _*+\ft14\omega _{[\mu }{}^{ab}(e,\psi )\gamma _{ab}+\ft12 \gamma _{[\mu }(\rmi\slashed{A}+\bar u)\right) \psi
_{\nu ]} \,, \nonumber\\
\hat{\phi }_a&= \ft12\gamma ^b\widehat{R}_{ba}(Q)+\ft1{12}\gamma _a\gamma ^{bc}\widehat{R}_{bc}(Q)=-\ft13\gamma ^b\widehat{R}_{ab}(Q)+\ft1{12}\gamma _{abc}\widehat{R}^{bc}(Q)\,,\nonumber\\
\widehat{\cal D}_a u&=\partial_a u+\bar{\psi}_a P_L\gamma\cdot\hat{\phi }\,,\nonumber\\
  \widehat{{\cal D}}_a A_b&=\nabla _aA_b+\ft12\rmi\bar \psi _a\gamma _* \hat{\phi }_b\,,\nonumber\\
\widehat{R}_{\mu \nu }{}^{ab}&= {R}_{\mu \nu }{}^{ab}-\bar \psi _{[\mu }\gamma ^{[a}\widehat{R}^{b]}{}_{\nu ]}(Q)+\ft12\bar \psi _{[\mu }\gamma_{\nu ]}\widehat{R}^{ab}(Q)\,,\nonumber\\
\widehat{R}_{ab}&=R_{(ab)}-\ft12\bar \psi^c \gamma _{(a} \widehat{R}_{b)c}(Q)+\ft12\bar \psi _{(a}\gamma ^c\widehat{R}_{b)c}(Q)\,,\nonumber\\
\widehat{R}&=R+\bar{\psi}_a\gamma _b\widehat{R}^{ab}(Q)\,,\qquad \widehat{G}_{ab}=\widehat{R}_{ab}-\ft12\eta _{ab}\widehat{R}\,.
\label{covRicci}
\end{align}
Note that the relation between $\hat{\phi }_a$ and $\widehat{R}_{\mu \nu  }(Q)$ is the same as between the superconformal quantities $\phi _a$ and ${R}_{\mu \nu  }(Q)$.
The covariantized field equations are then
\begin{align}
 \kappa  \Theta(X^0) &= -3\rmi\widehat{{\cal D}} _a A^a+\ft12 \widehat{R}+3A^a A_a\,,\nonumber\\
  \sqrt{2}\kappa \Theta(\Omega^0) &= \gamma ^{ab}\widehat{R}_{ab}(Q)\,,\nonumber\\
  \kappa \Theta (F^0)&=-3u\,,\nonumber\\
  \kappa ^2\Theta(A)_a&=6A_a\,,\nonumber\\
  \kappa ^2\Theta(\psi)_a &=-\ft12 \gamma_{abc}\widehat{R}^{bc}(Q)\,,\nonumber\\
  \kappa ^2 \Theta(e)_{ab}&=\widehat{G}_{ab}+6A_a A_b+3\eta_{ab}\left(u\bar{u}-A^c A_c\right)\,.
\label{Thetaconfgauge}
\end{align}
This gives the following components of the real multiplet
\begin{align}
\kappa ^2\mathcal{C}_{a}=&\;6A_a\,,\nonumber\\
\kappa^2\mathcal{Z}_{a}=&-6\hat{\phi }_a
\,,\nonumber\\
\kappa^2 \mathcal{H}_{a}=&-6\rmi\widehat{\mathcal{D}}_{a}\bar u\,,\nonumber\\
\kappa ^2\mathcal{B}_{ab}=&\;3\widehat{G}_{ab}-\eta_{ab}\widehat{G}_c{}^c+18 A_a A_b-3\eta_{ab}A_c A^c+3\eta_{ab}u\bar{u}-3\varepsilon_{abcd}\widehat{{\cal D}} ^{c}A^{d}\,,\nonumber\\
\kappa ^2P_R\Lambda_a=&2P_R\gamma^b
\left(\widehat{\cal D}_{[a}+\ft{3}{2}\rmi A_{[a}\right){\cal Z}_{b]}-\rmi P_R\gamma ^b\gamma_{ac}{\cal Z}_b A^c\,,\nonumber\\
\kappa ^2 D_{a}=&-12\widehat{\cal D}_{[b}\widehat{\cal D}^b A_{a]} -18\varepsilon _{abcd}A^b\widehat{\cal D}^c A^d -18 A_a A_b A^b+9 \rmi u \overset{\leftrightarrow}{\widehat{\cal D}}_{a}\bar u-36 A_au\,\bar u \nonumber\\
&-\ft14\rmi\bar{{\cal Z}}_a\gamma_*\gamma\cdot{\cal Z}-\ft13\rmi\overline{{\cal Z}}^b\gamma_*\gamma_{[b}{\cal Z}_{a]}+\rmi\bar{\widehat{R}}_{ab}(Q)\gamma_*{\cal Z}^b\,.
  \label{ETMGauge}
\end{align}
The result (\ref{ETMGauge}) gives all components of the multiplet
\begin{equation}
  {\cal E}_a= -\ft43{\cal C}_a\,.
 \label{EinC}
\end{equation}
Its linearized part is  the result in \cite[(2.13)]{Ferrara:1977mv}. The result is given also in \cite{Townsend:1979js,Ferrara:1988qx} with different definitions of the components, as usual when comparing conformal and Poincar\'{e} transformations \cite{Ferrara:1978jt,Kugo:1982cu}. We explain the relations in detail in Appendix \ref{ss:poincmultiplet}. The result in (\ref{ETMGauge}) is more elegant due to the conformal symmetry. See in particular the conformal form (\ref{ETM}), which is independent of the choice of auxiliary fields for supergravity.

When one investigates preservation of supersymmetry (four supercharges) in this conformal gauge, one should consider the transformations of the fermions in (\ref{vectorreal}), using the combination of the symmetries as in (\ref{delPoincare}). Especially the vanishing of $\delta P_{L}\mathcal{Z}_{a}$ gives the condition for preservation of supersymmetry when we just take the bosonic part. This is
\begin{equation}
  \ft{1}{2}P_{L}\left(\rmi\mathcal{H}_{a}-\gamma^{b}\mathcal{B}_{ba}-\rmi\slashed{\mathcal{D}}\mathcal{C}_{a}\right)\epsilon+\ft12\rmi P_L \left(-3{\cal C}_a+\gamma _{ab} {\cal C}^b\right)(\rmi\slashed{A}-\bar u)\epsilon=0\,.
 \label{delZa0}
\end{equation}
This equation can be decomposed to the conditions (2.11) in \cite{Festuccia:2011ws}.
E.g. the only terms proportional to $\gamma _{ab}\epsilon $ lead to $uA_a=0$, which is the first of (2.11) in \cite{Festuccia:2011ws}.  When these are already zero, the only term without a gamma matrix is ${\cal H}_a=0$, which is $\partial _a u=0$, which is the third of (2.11) in \cite{Festuccia:2011ws}. Therefore we now concentrate on terms proportional to $P_L \gamma ^b\epsilon $. These are
\begin{equation}
  0=-\ft12\kappa ^2\left.\mathcal{B}_{ba}\right|_{\rm bos}-\rmi \partial _b A_a + 12 A_a A_b - 3\eta _{ab}A_cA^c\,.
 \label{delZab0}
\end{equation}
The imaginary part leads to $\nabla _b A_a=0$, which is the second in \cite{Festuccia:2011ws}. This leaves then only the symmetric part of $\mathcal{B}_{ba}$, and the equation
\begin{align}
  0= B_{ab}\equiv &\kappa ^2\left.\mathcal{B}_{(ab)}\right|_{\rm bos}-24 A_a A_b +6 \eta _{ab}A_cA^c\nonumber\\
   = &3R_{ab}-\ft12\eta _{ab}R -6A_a A_b + 3 \eta _{ab}A_cA^c +3\eta _{ab}u\bar u\,.
 \label{vanishingB}
\end{align}
The expression $B_{ab}$ agrees with the component obtained in \cite{Townsend:1979js,Ferrara:1988qx}.
The difference between the conformal part $\kappa ^2\mathcal{B}_{(ab)}$ and this Poincar\'{e} expression $B_{ab}$ originates from the $S$-supersymmetry and depends on the conformal gauge fixing. This difference is traceless. The trace of $\mathcal{B}_{(ab)}$ or $B_{ab}$ is proportional to the real part of the last component of ${\cal R}/X^0$, which will be discussed in the next subsection.\footnote{Since this is a conformal chiral multiplet with $w=0$, the fermion has no $S$-supersymmetry transformations, and the auxiliary fields of the conformal and Poincar\'{e} multiplets coincide.}
To obtain the last of (2.11) in \cite{Festuccia:2011ws}, one considers
\begin{align}
  0= \ft13 B_{ab}+ \ft16 \eta _{ab}\eta ^{cd}B_{cd}
  =R_{ab}-2A_aA_b+2\eta _{ab}A^cA_c+3\eta _{ab}u\bar u\,.
 \label{Rabeqn}
\end{align}
To obtain the further condition of vanishing Weyl tensor in  \cite{Festuccia:2011ws}, one simply looks at the variation of the fermionic component $W_{\alpha\beta\gamma}$ (see \eqref{WRQ}, \eqref{varfermcomp}) of the superconformal Weyl multiplet. This gives the vanishing of equation \eqref{Ctensor}, which implies the vanishing of the Weyl tensor and of the $A_\mu$ field strength.
\medskip

The components of the current $J_\mu $ are obtained from (\ref{ETM}) by using the (covariantized) field equations of the other part of the action:
$[3X^0\bar X^{\bar 0}\Phi _{{\rm M}}(S,\bar S)]_D$.  For the conformal case, this can be written as $[\Phi _{{\rm M}}(X,\bar X)]$, where the $X$-dependence does not include $X^0$. In that case, the gauge conditions are not relevant in that part, and we can directly use (\ref{ETM}).

\subsection{The supersymmetric Ward identity}

Now that we have the full expressions of the components of ${\cal E}_a$, we can explicitly check (\ref{sconfeqnER}), which we write here again as
\begin{equation}
\overline{{\cal D}}^{\dot{\alpha}} {\cal E}_{\alpha\dot{\alpha}}=(X^0 )^3{\cal D}_\alpha \left(\frac{{\cal R}}{X^{0}}\right)\,.
\label{eq:Einstein equations}
\end{equation}
Since
\begin{equation}
  T(\bar X^{\bar 0})=\{\bar F^{\bar 0},\,\slashed{\cal D}P_R\Omega^{\bar 0},\,\bbox^C \bar X^{\bar 0}\}\,,
 \label{TX0comp}
\end{equation}
the components of the multiplet ${\cal R}/X^0$ are:
\begin{align}
  \frac{{\cal R}}{X^{0}}=\frac{T(\bar X^{\bar 0})}{(X^{0})^2}=& \left\{\frac{\bar{F}^{\bar0}}{(X^{0})^2},\, \frac{1}{(X^{0})^2}P_{L}\slashed{\mathcal{D}}\Omega^{0}-2\frac{\bar{F}^{\bar0}}{(X^0)^3}P_{L}\Omega^{0},\,\right.\nonumber\\
 & \left.\frac{1}{(X^{0})^2}\Box^{c}\bar{X}^{\bar0}
-2\frac{\bar{F}^{\bar0}}{(X^{0})^3}F^{0}+\frac{2}{(X^{0})^{3}}\bar{\Omega}^{0}P_{L}\slashed{\mathcal{D}}\Omega^{0}-3\frac{\bar{F}^{\bar0}}{(X^0)^{4}}\bar{\Omega}^{0}P_{L}\Omega^{0}\right\}\,.
 \label{calRoverX0}
\end{align}
This is a multiplet with Weyl weight~0, which therefore can be a constant without breaking supersymmetry, a possibility that we shall consider in Sec. \ref{ss:SCAdS}. The superconformal covariant derivatives are defined in (16.34) and (16.37) of \cite{Freedman:2012zz}. For convenience we repeat here the bosonic part:
\begin{equation}
 \left. \bbox^C X^0\right|_{\rm bos} = (\nabla ^a-2b^a -\rmi A^a)D_a X^0-\ft16 R\,X^0\,,\qquad D_aX^0 \equiv  e_a^\mu \left(\partial _\mu-b_\mu  -\rmi A_\mu\right)X^0\,.
 \label{boxCX0bos}
\end{equation}
Therefore, the bosonic part of the last component of ${\cal R}/X^0$ has the following real and imaginary parts
\begin{equation}
-\ft 16\kappa  B^{\,\,a}_a=-\ft 16 \kappa^3{\cal B}^{\,\,a}_a\,, \qquad \rmi \kappa  \nabla^a A_a\,.
\end{equation}

In order to evaluate the right-hand side of (\ref{eq:Einstein equations}) we write with (\ref{alphaalphadot}), (\ref{EinC}) and (\ref{vectorreal}):
\begin{align}
\overline{{\cal D}}^{\dot{\alpha}} {\cal E}_{\alpha\dot{\alpha}}&=\ft14 \rmi(\gamma^a)_{\alpha\dot{\alpha}}\overline{{\cal D}}^{\dot{\alpha}}{\cal E}_{a}
=-\ft13\rmi(\gamma^a)_{\alpha\dot{\alpha}}\overline{{\cal D}}^{\dot{\alpha}}{\cal C}_{a}=\ft13\rmi(\gamma^a)_{\alpha}{}^{\dot{\alpha}}\overline{{\cal D}}_{\dot{\alpha}}{\cal C}_{a}= \ft16\left(\gamma^{a}P_{R}{\cal Z}_{a}\right)_\alpha .
\label{DEinZ}
\end{align}
Using (\ref{ETM}) we find
\begin{equation}
 \gamma^{a}P_{R}{\cal Z}_{a} = -P_L\gamma^{a}\Theta(\psi ) _{a} + 2\sqrt{2}\Omega ^I\Theta (F)_I= \sqrt{2}\left[X^I\Theta (\Omega )_I +2 \Omega ^I\Theta (F)_I\right]\,,
 \label{gammatraceZ}
\end{equation}
where we used $W(S)=0$ from (\ref{WardIdentities}). Using now the specific case of pure supergravity with (\ref{Thetacompensating}), we thus have
\begin{align}
\ft16\gamma^{a}P_{R}{\cal Z}_{a}= \frac{1}{\sqrt{2}} P_L X^0\slashed{\cal D}\Omega ^0 -\sqrt{2}\bar{F}^{\bar 0} P_L \Omega ^0\,,
\label{X3dRX0}
\end{align}
For the right-hand side of \eqref{eq:Einstein equations} we use the fermionic component of (\ref{calRoverX0}), and the local version of (\ref{Dalphachiralm}), and find indeed the same expression.
We thus find that the derived component of the Einstein tensor multiplet correctly obeys the Einstein Ward Identity.

\subsection{The conformal case}
\label{ss:conformalcase}

We now restrict ourselves to the `conformal case', $\Delta K=\Delta W=0$. This case can also be characterized by the fact that in the basis of the chiral fields $\{X^I\}=\{X^0,\, X^i\}$, the action is completely separated in\footnote{To connect to much of the usual literature we often use the variables $S^i$ rather than $X^i$, but since the transformation (\ref{introdS0Si}) is invertible, this does not change the conclusions of this section.}
\begin{equation}
  {\cal S}= \left[-3X^0\bar X^{\bar 0}\right]_D +\left[3\Phi_{\rm M} (X^i,\bar X^{\ib})\right]_D+ \left[W(X^i)\right]_F\,,
 \label{Ssplit}
\end{equation}
where $\Phi_{\rm M}$ is homogeneous of first order in as well $X^i$ as $\bar X^{\ib}$ and $W$ of third order in $X^i$,
see (\ref{restrictPhiM}). We consider the conformal gauge fixing (\ref{gfconf}) where the transition to the super-Poincar\'{e} theory does not mix the two terms. Also we do not eliminate the auxiliary field $A_\mu $ of the Weyl multiplet, which is hidden in the notation in (\ref{Ssplit}), such that the splitting is preserved.

Therefore, $F^0$ appears only in the pure supergravity part (first term in (\ref{Ssplit})), and its field equation is $F^0\approx 0$. This is, up to invertible redefinitions, the vanishing of the first component of the chiral scalar curvature ${\cal R}$. It implies then the vanishing of all the components, i.e. the vanishing of (\ref{TX0comp}).
When we go to the conformal gauge, these equations reduce, using \cite[(16.42)]{Freedman:2012zz}, to
\begin{equation}
  \kappa \left.T(\bar X^{\bar 0})\right|_{\poinc}=\left\{u,\,\frac{\sqrt{2}}{6}\gamma \cdot \hat{\phi },\,-\rmi\hat{{\cal D}}{}^a A_a -\ft16 \hat{R}-A_aA^a\right\}\,.
 \label{confgaugecompcalR0}
\end{equation}
The bosonic part of these equations is discussed in detail in Sec. \ref{ss:CCJsugra}. Note that the fermionic part contains the $\gamma $-trace of the Rarita--Schwinger equation.

\subsection{Superconformal formulation of \texorpdfstring{$AdS_4$}{AdS} Supergravity}
\label{ss:SCAdS}
It is interesting to observe that $AdS_4$ \cite{Townsend:1977qa,Ferrara:1978rk} supergravity has a very simple description in our formalism. This theory corresponds to the
case when the superpotential is simply
\begin{equation}
{\cal W}=X_0^3\,.
\end{equation}
and no matter multiplets are present. This corresponds to scalar curvature (superconformal) multiplet ${\cal R}/X^0$ to have only a non-vanishing first component (see (\ref{Rfieldeq})), which gives now ${\cal R}\approx -2X^0$).  In particular, the vanishing of the last component, see (\ref{calRoverX0}), needed for supersymmetry being unbroken, gives in the conformal gauge (setting the fermions and $A_\mu$ to vanish)
\begin{equation}
\kappa ^{-1}\left. \frac{{\cal R}}{X^0}\right\vert_{\text{last}}=-\ft 16 R - 2u\,\bar u\,,\label{lastR}
\end{equation}
which vanishes for $R=-12 u\, \bar u$.  If one looks at the Einstein multiplet, the only possible non vanishing term is in the $\mathcal{B}_{ab}$ component
\begin{equation}
\kappa^2 \mathcal{B}_{ab}=3G_{ab}-\eta_{ab} G_c{}^c+3\eta_{ab}u\, \bar u\,. \label{kBab}
\end{equation}
However, since only the trace is possibly non vanishing we get
\begin{equation}
\kappa^2 \eta^{ab} \mathcal{B}_{ab} =- G_c{}^c+12u\, \bar u\,, \label{keab}
\end{equation}
but $G_c{}^c =-R$, so we obtain $R+12 u\, \bar u$ as the last component of ${\cal R}/X^0$ \eqref{lastR}.
This must in fact be the case if the identity
$\overline{{\cal D}}{}^{\dot \alpha }{\cal E}_{\alpha \dot \alpha }= (X^0)^3 {\cal D}_\alpha \left(\frac{{\cal R}}{X^0}\right)$ is satisfied.

The outcome therefore is that in the Minkowski and $AdS_4$ backgrounds the Einstein tensor ${\cal E}_{\alpha \dot \alpha } \approx 0$, but in $AdS_4$ is $\frac{{\cal R}}{X^0}=(\kappa u, 0, 0)$ with constant $u$. However, notice that the $AdS_4$  background can not be recovered in the linearized approximation   \cite{Ferrara:1977mv}, because of the nonlinear nature of equations \eqref{lastR}, \eqref{kBab}, \eqref{keab}.

\subsection{Superconformal formulation of \texorpdfstring{$dS_4$}{dS} Supergravity}
\label{dS}
The superconformal approach is also suitable to discuss de Sitter supergravity with a Volkov-Akulov chiral nilpotent superfield $X$ ($X^2=0$). In this case we can take the $\Phi$ function still conformal invariant
\begin{equation}
-\frac{N}{3}=X^0\bar X^{\bar 0}- X\bar X\,,
\end{equation}
and the superpotential \cite{Antoniadis:2014oya,Bergshoeff:2015tra,Hasegawa:2015bza}
\begin{equation}
{\cal W}(X)=\mu X (X^0)^2 + \lambda (X^0)^3 \, .
\end{equation}
For $\mu =0$ we get back to $AdS_4$ supergravity. We note that since $X^2=0$ the Poincar\'{e} and conformal gauges are the same, but the latter is simpler since the potential is given by  (\ref{nonconfV})
\begin{equation}
 \left. \kappa^{-4}V(S,\bar S)\right|_{S, \bar S=0}=\ft13\kappa^{-4}\left(\mu^2-9 \lambda ^2\right)\,.
\end{equation}
for $S=0$. Since $\mu \neq 0$ and $F^0$ is as before, the last component of ${\cal R}/X^0$ multiplet is  not vanishing, while the first component would vanish if $\lambda =0$. In an analogous way the
$ \mathcal{B}_{ab}$ component of the Einstein tensor would be non vanishing. However, due to the additive nature of the terms
\begin{align}
&F^0 \approx   \kappa^{-2} \lambda\,, \qquad F^1 \approx -\ft13 \kappa^{-2}\mu\, , \nonumber\\
& \kappa {\cal R}|_{\text{last}}=-\ft16 R \kappa -2 \kappa^3 |F_0|^2 \approx
 -\ft 29 \kappa^{-1} \mu^2\,, \qquad  \left.\kappa{\cal R}\right\vert_{\text{first}} \approx \lambda\,,
\end{align}
(using $R\approx 4\kappa ^{-2}V$)
as well the trace part of $ \mathcal{B}_{ab}$  will be
\begin{equation}
\kappa^2\mathcal{B}_{ab}\approx -\ft32 \eta_{ab}{\cal R}|_{\text{last}}=\ft13 \kappa ^{-2}\eta_{ab} \mu^2\,,
\end{equation}
and these formulae show that de Sitter supergravity breaks supersymmetry.

\subsection{Superconformal formulation of \texorpdfstring{$S^3\times L$}{S3 x L} supergravity}\label{S3timesL}
Another solution preserving full supersymmetry, considered in \cite{Festuccia:2011ws}, is the product of the 3-sphere and a line, obtained by taking
\begin{eqnarray}
A_a=(A_0, \, 0, \, 0, \, 0)\,, \qquad u=0\,, \qquad B_{ab}=0\,,
\end{eqnarray}
where $A_0$  is a constant. Then it is easy to see from (\ref{Rabeqn}) that the Ricci tensor is
\begin{eqnarray}
R_{00}=R_{0i}=0, \quad R_{ij}=2 A_0^2\delta_{ij}\,, \qquad (i=1,2,3)\,.
\end{eqnarray}
The space is conformally flat, so the Weyl tensor vanishes. In this background the two chiral multiplets (Weyl and scalar curvature) vanish $W_{\alpha\beta\gamma}={\cal R}/X^0=0$, while the Einstein tensor has one non-vanishing first component using the basis of Poincar\'{e} components of a real vector multiplet described in Appendix \ref{ss:poincmultiplet}:
\begin{equation}
E_a^{\rm P}=(-8A_a, \, 0,..., 0)\,.
\end{equation}
This space as well as $AdS$ satisfies $\overline{\cal D}^{\dot\alpha}E_{\alpha\dot\alpha}=(X^0)^3 {\cal D}_\alpha \left(\tfrac{{\cal R}}{X^0}\right)=0$.

\section{CCJ in supergravity}
\label{ss:CCJsugra}

In this section we consider the bosonic part of the action, which clarifies how our results modify (covariantize) the equations of CCJ \cite{Callan:1970ze}.
\subsection{Bosonic part of pure supergravity and currents}

We first consider the bosonic part of the superconformal version of pure supergravity:
\begin{align}
  {\cal S}_{\rm bos,SG}&= \left[-3X^0\bar X^0\right]_D=3\int \rmd^4x\,e\left[-F^0\bar F^0- \Re X^0\bbox^C \bar X^0 \right]\nonumber\\
  &=3 \int \rmd^4x e\left[-F^0\bar F^0 + D_aX^0\,D^a\bar X^0 +\ft16 R X^0\bar X^0\right]\,,
 \label{SbosSG}
\end{align}
where $D_aX^0$ is given in (\ref{boxCX0bos}).
This leads to the covariant field equations, as defined in (\ref{Thetadefs}),
\begin{eqnarray}
 \Theta(e) _{ab}& = & 3\eta _{ab}\left[F^0\bar F^0 - D_cX^0D^c\bar X^0 \right] \nonumber\\
   &   & +6 D_{(a} X^0\,D_{b)}\bar X^0 +\left(\eta _{ab}{\cal D}^cD_c-{\cal D}_{(a} D_{b)}\right) (X^0\bar X^0)\,,\nonumber\\
 \Theta(A) _a & = &-3\rmi X^0D_a\bar X^0 +3\rmi \bar X^0D_a X^0 \,,\nonumber\\
  \Theta (X)_0&=&-3{\cal D}^aD_a \bar X^0= -3D^a D_a \bar X^0 +\ft12 R \bar X^0 \,,\nonumber\\
\Theta (F)_0&=&-3\bar F^0\,,
 \label{fieldeqns}
\end{eqnarray}
where $D_a$ is covariant for the linearized symmetries, but ${\cal D}_a$ is the fully conformal covariant derivative, which makes a difference for
\begin{equation}
  {\cal D}_{(a} D_{b)} (X^0\bar X^0) = \left(\nabla _a\partial _b+4 f_{(ab)}\right)(X^0\bar X^0)= \left(\nabla _{(a}\partial _{b)}-R_{ab}+\ft16 \eta _{ab}R \right)(X^0\bar X^0)\,,
 \label{DDXbarX}
\end{equation}
using (\ref{depfields}).  The latter terms produce in $\Theta (e) _{ab}$ the term $G_{ab}X^0\bar X^0$.

We now consider these equations in the conformal Poincar\'{e} gauge (\ref{gfconf}). Defining $u=\kappa \bar F^0$, the bosonic part of the pure supergravity action is
\begin{equation}
 \left. {\cal S}_{\rm bos,SG}\right|_{\poinc}= \int \rmd^4x\,e \kappa ^{-2}\left[\ft12 R-3 u\bar u +3A_\mu A^\mu  \right]\,.
 \label{SSGbosPoinc}
\end{equation}
The field equations in (\ref{fieldeqns}) are
\begin{align}
 \kappa ^2 \Theta(e) _{ab}  =&G_{ab}-L_{ab}\,,\qquad L_{ab}=-3\eta _{ab}u\bar u
   -6A_a A_b +3\eta _{ab}A_cA^c  \nonumber\\
  \kappa ^{2}\Theta _a(A) = & 6A_a\,,\qquad \kappa\Theta (X)_0= 3 A_aA^a+\ft12 R-3\rmi \nabla ^aA_a\,,\qquad \Theta (F)_0=-3\bar F^0\,.
\label{ThetaSGPoinc}
\end{align}
Since ${\cal D}_aX^0= -\rmi A_a\kappa ^{-1}$, the remaining terms of the Ward identity in (\ref{WardIdentities}) are
\begin{equation}
 \kappa ^2 \left.W(P)_a\right|_{\poinc}=\nabla ^{b}\Theta (e)_{ab}+6A^b\left(\partial _a A_b-\partial _b A_a\right)-6A_a\nabla ^bA_b-3\partial _a(u\bar u)=0\,.
 \label{WPPoinc}
\end{equation}
Due to the Bianchi identity (\ref{GRanalogue}), $D^bG_{ab}=0$, we get for $L_{ab}$:
\begin{equation}
  \nabla ^bL_{ab}= 6A^b\left(\partial _a A_b-\partial _b A_a\right)-6A_a\nabla ^bA_b-3\partial _a(u\bar u)\,.
 \label{DbLab}
\end{equation}

Adding the matter Lagrangian (without eliminating auxiliary fields), and defining
\begin{equation}
  \Theta^c_{ab}=e^{-1}e_b^\nu \frac{\delta {\cal S}_{\rm matter}}{\delta e^{\nu a}}=2e^{-1}e_a^\mu e_b^\nu \frac{\delta {\cal S}_{\rm matter}}{\delta g^{\mu\nu}}\,,
 \label{ThetacabBos}
\end{equation}
the gravitational field equation is
\begin{equation}
  G_{ab }-L_{ab }+\Theta^c_{ab}\approx 0\,,
 \label{gravFEmatterBos}
\end{equation}
and the conservation equation due to the same Bianchi identity is
\begin{equation}
 \nabla ^b\Theta^c_{ab}\approx  \nabla ^b L_{ab}\,.
 \label{conservThetac}
\end{equation}
The right-hand side is given by (\ref{DbLab}). In case of conformal coupling, $u\approx 0$ and $\nabla ^a A_a\approx 0$ (see (\ref{calR0bosonic})) and
\begin{equation}
  \nabla ^b\Theta^c_{ab}\approx 6A^b\left(\partial _a A_b-\partial _b A_a\right)\,.
 \label{consThetacconf}
\end{equation}
This shows the modification of the conservation equation for the conformal case. In that case, we could also consider the trace condition. Since
\begin{equation}
  L_a{}^a = 6A_aA^a -12 u\bar u\,,
 \label{Laa}
\end{equation}
the tracelessness  of the improved energy-momentum tensor for the conformal case ($u\approx 0$) follows from
\begin{equation}
  \Theta^c{}_a{}^a \approx  R+ 6A_aA^a\approx 0\,.
 \label{Thetacaa}
\end{equation}

The above equations are in agreement with Sec. \ref{ss:conformalcase}, where we found that in the conformal case we obtain ${\cal R}\approx 0$, i.e. $T(X^0)$, which in conformal gauge lead to the component expressions (\ref{confgaugecompcalR0}). The  bosonic part is of the latter is
\begin{equation}
  u^0\equiv \kappa \bar F^0\approx 0\,,\qquad \nabla^\mu A_\mu \approx 0 \,,\qquad  R +6A_\mu^2\approx 0\,.
 \label{F0components}
\end{equation}
\subsection{Bosonic action and improved currents}
We can write down the action of the matter-sugra coupled action from \cite[(17.19)]{Freedman:2012zz} with
\begin{equation}
  N=3 X^0\bar X^{\bar 0}\left(-1+\Phi _{{\rm M}}(S,\bar S)\right)\,, \qquad {\cal W}=(X^0)^3 W(S)\,.
\label{NWgeneral}
\end{equation}
In the \emph{conformal case}, to which we will first restrict ourselves, the homogeneity of $\Phi _{\rm M}$  and $W$ allow us also to write
\begin{equation}
  \ft13N=   -X^0\bar X^{\bar 0}+\Phi _{{\rm M}}(X,\bar X)\,,\qquad {\cal W}= W(X)\,,
\label{NWconformal}
\end{equation}
where the dependence on $X$ and $\bar X$ is restricted to dependence on the $X^i$ and their complex conjugates. Therefore, only the first term will lead to a breaking of superconformal symmetry to super-Poincar\'{e}.

The bosonic part of the action is then
\begin{align}
{\cal S}=\int \rmd^4x\,\sqrt{-g}& \left[ \ft12 R X^0\bar X^{\bar 0}+3D_\mu X^0D^\mu \bar X^{\bar 0} - 3F^0\bar F^{\bar 0}\right.\nonumber\\
&- \ft12 R \, \Phi _{\rm M}(X,\bar X)
+ 3\Phi _{{\rm M}\,i\jb}(X,\bar X)\left( -D_\mu X^iD ^\mu \bar X^{\jb}+ F^i\bar F^{\jb}\right)\nonumber\\
&\left.+ F^i W_i(X) +\bar F^{\ib} W_{\ib}(\bar X)
\right]\,,\nonumber\\
&D_\mu X^I\equiv (\partial _\mu -\rmi A_\mu )X^I\,.
 \label{action1}
\end{align}
The first line is the pure supergravity action. The second and third line do not depend on $X^0$. They are separately conformal invariant and this will not be broken in the conformal gauge.
 The first equation in (\ref{F0components}) is obvious from this form of the action.

After fixing the gauge $X^0=\bar X^{\bar 0} =\kappa^{-1}$ and taking into account that $X^i=X^0 S^i =\kappa^{-1} S^i$, the action becomes
\begin{align}
{\cal S} =&\int \rmd^4x\,\ft12 \sqrt{-g}\left[ \kappa^{-2} \left( R +6 A_\mu^2-6\Phi_{{\rm M} i\jb}  g^{\mu\nu} D_\mu S^i D_\nu \bar S^{\jb}-R \Phi_{\rm M}(S,\bar S) \right) -2\kappa^{-4} V(S,\bar S) \right]
\,,  \label{lagAmu}
\end{align}
where here and below $\Phi_{\rm M}$ is considered as function of $S^i$ and $\bar S^i$.
The term $-R \Phi_{\rm M}$ completes the conformal coupling and gauge invariant coupling and the covariant derivatives are
\begin{equation}
D_\mu S^i =(\partial_\mu -\rmi A_\mu) S^i\,,\qquad D_\mu \bar S^{\ib} =(\partial_\mu +\rmi A_\mu) \bar S^{\ib}\,.
 \label{Dmuz}
\end{equation}

The $F^i$ have been integrated out, and produced a potential
\begin{equation}
 V(S,\bar S)=\ft13  (\Phi _{{\rm M}\,i\jb})^{-1} W_i(S) W_{\jb}(\bar S)\,,
 \label{VSbarS}
\end{equation}
which is homogeneous of second degree in $S$ and in $\bar S$.

We will below also consider a potential that is not conformal in order to allow mass terms. In that case $W$ is not homogeneous of third order in $S^i$, but there is still the conformal
${\cal W}(X)= (X^0)^3 W(X^i/X^0)$ as in (\ref{actS}).\footnote{In view of the normalisation of the kinetic terms of the scalar fields, the physical fields in the conformal gauge are $X^i$ rather than $S^i$, then setting $S^i=\kappa X^i$ in our formulae we see that the superpotential ${\cal W}=\kappa ^{-3}W(\kappa X)$, of dimension 3 gets dimensionful coefficients
respectively of dimension $3, 2,1$ for the constant, linear and quadratic terms in $X$, while the cubic term remains dimensionless.} The third line of (\ref{action1}) is then
\begin{align}
 F^I {\cal W}_I +\hc =& (X^0)^2\left[F^0 (3\Delta W) + F^i W_i (S)\right] +\hc \,,\qquad 3\Delta W =3 W(S)  - S^i W_i(S)\,.
 \label{nonconfFterms}
\end{align}
The elimination of the auxiliary $F$-terms then leads to a potential (in conformal gauge) that is a generalization of (\ref{VSbarS}):
\begin{equation}
  V(S,\bar S)=\ft13\left((\Phi _{{\rm M}\,i\jb})^{-1} W_i(S) W_{\jb}(\bar S) -  |3 \Delta W|^2\right)\,.
 \label{nonconfV}
\end{equation}
In the Einstein gauge, (\ref{gfN}), we get
\begin{eqnarray}
V^{\rm P}=\frac{1}{(1-\Phi _{\rm M})^2} V.
\end{eqnarray}
This is in agreement with the direct calculation of the potential of Poincar\'{e} supergravity using K\"{a}hler geometry.

The simplest kinetic terms appear for $\Phi_{\rm M}=S ^i \bar S^ {\ib}$, and corresponds thus to $\Phi_{{\rm M}i\jb}=\delta _{i\jb}$. This is the $\mathbb{C}\mathbb{P}^n$ model.

Now we proceed as CCJ, namely we write the trace of the Einstein equations and the scalar field equations (working with $A_\mu$ off-shell).  The Einstein equations coming from the Lagrangian \eqref{lagAmu} plus the CCJ improvement term has the following form
\begin{align}
&R_{\mu\nu}-\ft12 R g_{\mu\nu} + 6 A_\mu A_\nu -3 g_{\mu\nu} A_\rho^2  -(R_{\mu\nu}-\ft12 R g_{\mu\nu})\Phi_{\rm M}+ (\nabla _\mu \partial_\nu - g_{\mu\nu} \nabla ^2)\Phi_{\rm M}\nonumber\\
&-3\Phi _{{\rm M}\,i\jb}(D_\mu S^i D_\nu \bar S^{\jb} +D_\mu \bar S^{\jb} D_\nu S^i -g_{\mu\nu} D_\lambda S^i D^\lambda \bar S^{\jb})+g_{\mu\nu} \kappa^{-2}V(S^i,\bar S^{\ib}))
\approx 0\,.\label{Eqccj}
\end{align}
Taking the trace, $g^{\mu\nu}\frac{\delta {\cal S}}{\delta g^{\mu\nu}}=0$, using the homogeneity of $\Phi_{\rm M} $, one gets
\begin{equation}
-(R+6 A_\mu ^2)+ R \Phi_{\rm M}  -3 \left(\Phi _{{\rm M}\,\jb}D^2 \bar S^{\jb}+ \Phi _{{\rm M}\,i} D^2 S^i\right)+ 4 \kappa^{-2}V(S^i,\bar S^{\ib})\approx 0\,. \label{trwithA}
\end{equation}
Now we can use the $S^i$ and $\bar S^{\ib}$ field equations in the last term, which are
\begin{eqnarray}
3D_\mu D^\mu  S^i\approx \ft12 R\,S^i + \kappa^{-2} (\Phi _{{\rm M}\,i\jb})^{-1} V_{\jb}\,,\label{compeomz}\\
3D_\mu D^\mu \bar S^{\jb}\approx \ft12\,R \bar S^{\jb}+ \kappa^{-2} (\Phi _{{\rm M}\,i\jb})^{-1}  V_i\,,   \label{compeomzbar}
\end{eqnarray}
where the covariant derivatives are $\U(1)$ and general covariant:
\begin{eqnarray}
D_\mu D^\mu S^i =\frac{1}{\sqrt{-g}}(\partial_\mu -\rmi A_\mu )\sqrt{-g}(\partial^\mu -\rmi A^\mu )S^i\,.
\end{eqnarray}
Multiplying \eqref{compeomz} by $\Phi _{{\rm M}\,i}$ and \eqref{compeomzbar} $\Phi _{{\rm M}\,\jb}$, we find
\begin{eqnarray}
3\Phi _{{\rm M}\,i} D_\mu D^\mu S^i\approx \ft{1}{2}R\, \Phi_{\rm M} + \kappa^{-2} \bar S^{\ib}V_{\ib}\,,\nonumber\\
3\Phi _{{\rm M}\,\ib}  D_\mu D^\mu \bar S^{\ib}\approx \ft 12 R\,\Phi_{\rm M}+ \kappa^{-2} S^i V_i\,. \label{zeqzbar}
\end{eqnarray}
Inserting these in \eqref{trwithA} we find
\begin{equation}
-(R+6 A_\mu ^2)\approx \kappa^{-2}\Delta V\,,\qquad \Delta V\equiv 4 V(S^i,\bar S^{\ib})-\bar S^{\ib}  V_{\ib}-  S^i  V_i\,.
\label{traceeom}
\end{equation}
This means that we find the last of (\ref{F0components})
\begin{equation}
R+6 A_\mu ^2\approx 0\,, \label{RplusA20}
\end{equation}
if $\Delta V$ vanishes, that is if $V(S^i,\bar S^{\ib})$ is homogeneous of degree 4 in $S^i$ and $\bar S^{\ib}$, i.e. if the holomorphic superpotential is homogeneous of degree of 3.

We can find the modified $\Theta _{\mu\nu}^c$ from \eqref{Eqccj}
\begin{eqnarray}
R_{\mu\nu}-\ft12 g_{\mu\nu} R + 6 A_{\mu} A_{\nu}-3g_{\mu\nu}A_\rho ^2 & \approx -\Theta _{\mu\nu}^c
\label{ModEinsteinThetac}
\end{eqnarray}
which can be written as
\begin{align}
\Theta _{\mu\nu}^c=&-3\Phi _{{\rm M}\,i\jb}\left(D_\mu S^i D_\nu \bar S^{\jb} +D_\mu \bar S^{\jb} D_\nu S^i- g_{\mu\nu}D^\lambda S^i D_\lambda \bar S^{\jb}   \right)+\left(\nabla  _\mu \partial_\nu -g_{\mu\nu} \nabla ^2  \right) \Phi_{\rm M}\nonumber\\
& - G_{\mu\nu} \Phi_{\rm M}+ g_{\mu\nu}\kappa^{-2} V(S^i, \bar S^{\ib})\,, \label{imprEMV}
\end{align}
with the property that
\begin{equation}
\Theta_\lambda ^{c \, \, \lambda}\approx R \Phi_{\rm M}-3\Phi _{{\rm M}\,\ib} D^2 \bar S^{\ib}- 3\Phi _{{\rm M}\,i}D^2 S^i + 4 \kappa^{-2} V (S^i, \bar S^{\ib})\approx 0\,,
\end{equation}
consistent with  (\ref{ModEinsteinThetac}) and (\ref{RplusA20}). The divergence of $\Theta _{\mu\nu}^c$ gives \eqref{consThetacconf}, so that bringing the $A$ terms on the right-hand side we have
\begin{equation}
G_{\mu\nu}\approx -\Theta _{\mu\nu}^c-6 A_\mu A_\nu+ 3g_{\mu\nu} A_\rho^2\,.
\end{equation}
Taking the divergence we have
\begin{equation}
\nabla ^\mu \Theta _{\mu\nu}^c+6 A^\mu (\partial_\mu A_\nu-\partial_\nu A_\mu)\approx0\,,
\end{equation}
since $\nabla^\mu G_{\mu\nu}=0$. To understand how the last equation follows, we obtain that the divergence of the matter part of the energy momentum tensor in \eqref{imprEMV} cancels the $A$ terms and the $\ft 12 R \partial_\nu \Phi_M$ of the $-G_{\mu\nu}\partial^\mu \Phi_M$ term, while the improvement term cancels the $-R_{\mu\nu}\partial^\mu \Phi_M$ part of the $-G_{\mu\nu}\partial^\mu \Phi_M$ term.

The expression (\ref{imprEMV}) is thus the improved energy-momentum tensor of CCJ that is improved by two types of covariantization. First, the derivatives are covariant for the $\U(1)$ $R$-symmetry. Furthermore there is the term $G_{\mu\nu} \Phi_{\rm M}$, which could be included in a conformal covariant derivatives:\footnote{$4 f_{\mu \nu}=-R_{\mu\nu}+\ft 16 g_{\mu\nu} R$ is given in \cite[(15.25)]{Freedman:2012zz}}
\begin{equation}
  {\cal D}_{(\mu}{\cal D}_{\nu )}\Phi _{{\rm M}}=  (\nabla  _\mu \partial_\nu + 4 f_{(\mu \nu )})\Phi_{{\rm M}} = (\nabla  _\mu \partial_\nu - R_{\mu \nu }+\ft16g_{\mu \nu })R\,,
 \label{DmuconfPhiM}
\end{equation}
which thus covariantizes the CCJ term
\begin{equation}
  \left({\cal D}  _{(\mu} {\cal D}_{\nu)} -g_{\mu\nu} {\cal D} ^2  \right) \Phi_{\rm M}=\left(\nabla  _\mu \partial_\nu -g_{\mu\nu} \nabla ^2  \right) \Phi_{\rm M}-G_{\mu \nu }\Phi_{\rm M}\,.
 \label{confcovCCJ}
\end{equation}

The  second of (\ref{F0components})  fixes the equation of motion of $A_\mu$. Let us take as in (\ref{EJPhi})
\begin{equation}
J_\mu =4\rmi \kappa^{-2}(\Phi _{{\rm M}\,i}D_\mu S^i- \Phi _{{\rm M}\,\ib} D_\mu \bar S^{\ib})\,.
\end{equation}
In the difference between the equations (\ref{zeqzbar}) in the conformal case the homogeneity properties of the potential imply the cancelation of the terms depending on $V$ and
\begin{equation}
 \nabla ^\mu J_\mu=4\rmi \kappa^{-2} (\Phi _{{\rm M}\,i} D^2 S^i-\Phi _{{\rm M}\,\ib}D^2 \bar S^{\ib}) \approx  0\,.
\end{equation}
The equation of motion for $A_\mu$ is
\begin{equation}
8 A_\mu\approx \kappa^2J_\mu \,, \label{AJmu5}
\end{equation}
in agreement with \eqref{EisJ} using (\ref{EaPoinc})  and gives
\begin{equation}
A_\mu\approx \frac{\rmi}{2(1- \Phi_{\rm M})}\Phi _{{\rm M}\,i\jb}(\bar S^{\jb}\partial_\mu S^i-S^i\partial_\mu \bar S^{\jb})
\end{equation}
and implying
\begin{equation}
2 \nabla  ^\mu A_\mu\approx \nabla  ^\mu J_\mu ^{5}\approx 0\,.
\end{equation}
This is the second of (\ref{F0components}).

\subsection{Deformation of a conformal potential and no-scale models}\label{DeformVnoscale}
We first discuss a particular deviation from conformal symmetry with just one complex scalar $S$, and we take the choices
\begin{equation}
\Phi _{\rm M}= S\bar S\,,\qquad W=\ft12(\lambda +S)^3\,.
\end{equation}
In this case we find that conformal invariance is broken by
\begin{equation}
\lambda W_S =3 W- S W_S=3 \Delta W\,,
\end{equation}
and the potential \eqref{nonconfV} becomes
\begin{equation}
V(S,\bar S)=\ft13  W_S W_{\bar S} \left(-|\lambda|^2+ 1\right)\,.
 \label{Vnonconf}
\end{equation}
This potential interpolates from $V$ positive to $V$ negative, where $\lambda=0$ corresponds to the conformal case. On the other hand for $|\lambda| =1$ the potential identically vanishes and this gives the single field example of no-scale supergravity \cite{Cremmer:1983bf} .

\subsection{CCJ and the equivalence principle}\label{CCJandequivalence}
In this subsection we discuss the equivalence principle following CCJ for conformally coupled gravity with a potential that breaks conformal invariance by a mass term. By inserting equation \eqref{traceeom} in \eqref{compeomz}, we obtain the matter field equation in the Einstein gauge.
\begin{eqnarray}
(D_\mu D^\mu  + A_\mu A^\mu )S^i \approx  \ft 16 \kappa^{-2}\left(\Delta V S^i +2 (\Phi_{\rm M i\jb})^{-1}V_{\jb}\right)\,.  \label{eomzDeltaV}
\end{eqnarray}
These equations are equivalent, up to field redefinition, to the standard  supergravity formulation of \cite{Cremmer:1982en}.
We consider minimal kinetic couplings and the mass $m$ being generated from an holomorphic superpotential of the form
\begin{eqnarray}
\Phi _{\rm M}=\delta _{i\ib} S^i \bar S^{\ib}\,,\qquad W(S)=\tfrac{m}{2} S^i S^i + \ft 13 \lambda_{ijk} S^i S^j S^k\,.
\end{eqnarray}
Since $\Delta V$ is at least quadratic in $S$, it means that the mass is not affected by gravitational interactions, while the interaction strengths are.
Looking at the form of the supergravity potential  \eqref{nonconfV}, we have at the quadratic order
\begin{eqnarray}
V=\tfrac{1}{3} m^2 S^i \bar S^{\ib}+ \text{higher order terms}\,,
\end{eqnarray}
so that the mass is not affected by the gravitational modification related to $\Delta V$, while higher interaction terms are. As anticipated by CCJ,  conformally coupled supergravity is then in agreement with the equivalence principle.

\section{Summary and conclusions}
\label{ss:summary}

The Einstein equations for matter-coupled supergravity in terms of the conformal tensor calculus have been obtained. We paid special attention to what we called 'the conformal case'. This is the supergravity coupling of ${\cal N}=1$ rigid supersymmetric models of chiral multiplets with conformal symmetry. In this case the K\"{a}hler couplings imply that there is a $\U(1)$ isometry group (the $R$-symmetry).\footnote{In the simplest case of the $\sigma$-model (\ref{sigmamod}) there is an additional $\SU(N)$ symmetry, which is not present in the other models satisfying the conformal restriction (\ref{restrictPhiM}).}

It has been relevant to consider the difference between two gauge choices for dilatations, which correspond to Einstein gauge and `conformal gauge'.
Going to the Einstein gauge  the scalar fields parametrize a K\"{a}hler $\sigma$-model with K\"{a}hler potential $K(S,\bar S)=-3 \log(-\Phi(S,\bar S)/3)$. The conformal case is  characterized by a homogeneity of $\Phi(S,\bar S)+3$ of order 1 both in $S$ and $\bar S$, and of the superpotential $W(S)$, which should be homogenous order $3$. However, in this gauge the conformal properties of the currents are not evident.

A conformal gauge preserves the separation between the pure supergravity part, where the superconformal symmetry is broken in order to get super-Poincar\'{e} gravity, and the matter part with preserved conformal symmetry. This separation is maintained by not eliminating the auxiliary gauge field $A_\mu $ of the  $\U(1)$ $R$-symmetry. The matter part has then still K\"{a}hler couplings, where now the K\"{a}hler potential is $\Phi(S,\bar S)$. The results provide a supersymmetric generalization of the properties of scalar fields coupled to gravity with improvement terms in CCJ \cite{Callan:1970ze}.
Two sort of bosonic improvement terms emerge, one that couples the scalar fields to the scalar curvature $R$, the other that couples the scalar fields to an $R$-current.
Both are part of the superconformal covariant derivatives that covariantize the (rigid conformal) CCJ improvement terms.
Therefore, the improved energy-momentum tensor that is traceless for superconformal matter contains also $\U(1)$ corrections. This also implies an improvement term in the $\U(1)$ current. These are part of the supercurrent, which becomes $\gamma$-traceless in the superconformal case \cite{Ferrara:1975pz} for which the compensator equation becomes the chiral superfield equation ${\cal R}\approx 0$ in (\ref{Rfieldeq}). We clarified the bosonic aspects separately in Sec. \ref{ss:CCJsugra}, which provides the improved currents for conformal K\"{a}hler couplings.

We have given explicit formulae, in the superconformal approach, for the three basic multiplets that specify the superspace geometry of ${\cal N}=1$ supersymmetry. These multiplets play a key role in the construction of higher curvature invariants and they have found applications to classify counterterms \cite{Stelle:2012zz,Bern:2014dlu}. More recently they were also relevant in cosmology to provide a generalization of the Starobinsky model as well as for nonlinear realizations for local supersymmetry in the framework of ${\cal N}=1$ supergravity \cite{Antoniadis:2014oya,Ferrara:2014kva,Ferrara:2015cwa,Bergshoeff:2016psz,Ferrara:2016ajl}. The latter is a particular way for implementing the super-Brout-Englert-Higgs effect and to find de Sitter vacua in cosmological scenarios. It is likely that our results will find new applications along this area of research.

Our results can also be relevant in exploring the interplay between different supergravity backgrounds, in the study of rigid supersymmetry in curved space. The simplest examples, preserving four supersymmetries were discussed in subsection \ref{ss:SCAdS} and \ref{S3timesL} and correspond to the conformally flat spaces $AdS_4$ and $S^3\times L$. Similar arguments in section \ref{dS} show that the $dS$ background is not supersymmetric.  Another related topic is the application of localization techniques in supersymmetric quantum field theories \cite{Festuccia:2011ws, Dumitrescu:2012ha, Pestun:2016zxk, Pufu:2016zxm}.

\medskip
\section*{Acknowledgments.}

\noindent
We thank Thomas Dumitrescu for discussions.
We acknowledge hospitality of the GGI institute in Firenze, where part of this work was performed during the workshop `Supergravity: what next?'. We also acknowledge hospitality from CERN during the string winter school 2017, from the Arnold-Regge center in Torino and from Mainz Institute for Theoretical Physics (MITP).
This work is supported in part by the COST Action MP1210 `The
String Theory Universe'.

The work of S.F. is supported in part by CERN TH Dept and INFN-CSN4-GSS.
The work of M.T. and A.V.P. is supported in part by the Interuniversity Attraction Poles Programme
initiated by the Belgian Science Policy (P7/37) and in part by
support from the KU Leuven C1 grant ZKD1118 C16/16/005. The work of M.T. is supported by the FWO odysseus grant G.0.E52.14N.

\bigskip


\appendix
\section{From components to superspace } \label{FromComptoSuperspace}
\label{app:superspace}

We translate components in superspace notation as in \cite[Appendix 14.A]{Freedman:2012zz}.
Thus e.g. the supersymmetry transformations of any quantity $X$ is
\begin{equation}
\delta\left. X\right|_{\theta =0} =\bar \epsilon^\alpha  P_L(D_\alpha X)_{\theta =0} + \bar \epsilon^{\dot\alpha } P_R (\bar D_{\dot \alpha }X)_{\theta =0}  \,.
 \label{transfotoD}
\end{equation}
The full 4-component spinor index is thus split in $(\alpha \dot \alpha )$ where the $\alpha $ part refers to the left-projected spinor, and $\dot \alpha $ to the right-projected spinor.

These notations for the chiral multiplet $S=\{Z,\,P_L\chi ,\, F\}$ imply (omitting the $\theta =0$ projection each time)
\begin{align}
  D_\alpha S &= \frac{1}{\sqrt{2}}(P_L\chi)_\alpha  \,, \qquad \bar D_{\dot \alpha} \bar S = \frac{1}{\sqrt{2}}(P_R\chi)_{\dot \alpha}\,,\qquad D^2 S = F\,.
  \label{Dalphachiralm}
\end{align}
We introduce here the notation $D^2$, which corresponds to
\begin{equation}
  D^2 = -\bar D P_L D \,,\qquad \bar D^2 = -\bar D P_R D\,,
 \label{D2barD2}
\end{equation}
where the bar in the right-hand sides is the Majorana bar. In 2-component notations these correspond to
\begin{equation}
  D^2 = -D^\alpha D_\alpha =D_\alpha  D^\alpha \,,\qquad \bar D^2= -\bar D^{\dot\alpha} \bar D_{\dot\alpha} =\bar D_{\dot\alpha}  \bar D^{\dot\alpha}
 \label{D22comp}
\end{equation}
It acts e.g. as
\begin{equation}
 \left. D^2 \bar \theta P_L\theta \right|_{\theta =0}= 4\,.   
 \label{D2theta2}
\end{equation}

We use also the operation $T$ on a multiplet which leads to a chiral multiplet. On the antichiral multiplet  $T(\bar S)$ is defined if $S$ has Weyl weight~1, and its first component is then $\bar F$. Its definition on other multiplets is defined in \cite{Ferrara:2016een} following the pioneering work in \cite{Kugo:1983mv}. In flat space $T=\bar D^2$.

We also use bispinor notation $V_{\alpha \dot \alpha }$ for vectors $V_\mu $ using the rule
\begin{equation}
  V_{\alpha \dot \alpha }=\ft14\rmi \gamma ^\mu _{\alpha \dot \alpha }V_\mu \,,\qquad V_\mu  = 2\rmi V _{\alpha \dot \alpha }(\gamma _\mu)^{\alpha \dot \alpha}\,.
 \label{alphaalphadot}
\end{equation}
In particular, this implies
\begin{equation}
  \partial_{\alpha \dot \alpha }=\ft14 \rmi\gamma ^\mu _{\alpha \dot \alpha }\partial _\mu= -\ft12\rmi
  \left(D_\alpha \bar D_{\dot \alpha }+ \bar D_{\dot \alpha}D_\alpha\right)\,,\qquad D_\alpha \bar D_{\dot \alpha }+ \bar D_{\dot \alpha}D_\alpha= 2\rmi \partial_{\alpha \dot \alpha }=-\ft12\gamma ^\mu _{\alpha \dot \alpha }\partial _\mu\,.
 \label{Dalgabra}
\end{equation}

A Fierzing leads e.g. to
\begin{equation}
  \ft14 \gamma ^\mu _{\alpha \dot \alpha }\bar \chi P_R\gamma _\mu \chi = -(D_\alpha S)(\bar D_{\dot \alpha }\bar S)\,.
 \label{Fierzresult}
\end{equation}

For $D$ and $F$-type actions we use the notation (for a real multiplet $C$ whose last component is $D$ and a chiral multiplet
\begin{equation}
 [C]_D  = \frac12 \int \rmd^4x\,e\left[D +\ldots  \right]\,,\qquad
  [X]_F = \int \rmd^4x\,e\, 2 \Re F+\ldots \,,
\label{DFactions}
\end{equation}
where the extra terms are determined by conformal invariance, and contain the fields of the Weyl multiplet $\{e_\mu ^a,\,\psi _\mu, \, b_\mu ,\,A_\mu \}$. In rigid supersymmetry in flat space, they correspond to the superspace expressions (identifying superfields by their first components)
\begin{equation}
  [C]_D =\int \rmd^4 x\,\rmd^4 \theta C=\int \rmd^4x\, D^2 \bar D^2 C\,, \qquad
  [X]_F= \int \rmd^4 x\,\rmd^2 \theta\, X +\hc=\int \rmd^4x\, D^2 X +\hc\,.
 \label{DFinttheta}
\end{equation}
\section{Weyl multiplet and constraints} \label{sec:constraints}
A priori the superconformal algebra is gauged by adding a gauge field for every generator in the algebra.  Constraints on some curvatures are imposed
\begin{align}
  0&=R_{\mu\nu}(P^{a})\,,\nonumber\\
  0&=\gamma^\mu R_{\mu\nu}(Q)\, ,\nonumber\\
  0&=e^\rho_b R^{\rm cov}_{\mu\rho}(M^{ab})-\rmi \tilde{R}_\mu{}^a (T)\,.  \label{eq:constraints}
\end{align}
where $R^{\rm cov}_{\mu\rho}(M^{ab})$ is the covariantized curvature of Lorentz transformations. These determine the gauge fields of local Lorentz rotations ($\omega _\mu {}^{ab}$), $S$-supersymmetry $(\phi _\mu )$ and special conformal symmetry ($f_\mu {}^a$)
\begin{align}
\omega _\mu {}^{ab}=&\;2 e^{\nu[a} \partial_{[\mu} e_{\nu]}{}^{b]} -
e^{\nu[a}e^{b]\sigma} e_{\mu c} \partial_\nu e_\sigma{}^c+2e_\mu {}^{[a}e^{b]\nu }b_\nu+\ft12\bar \psi
  _\mu \gamma ^{[a}\psi ^{b]}+\ft14\bar \psi ^a\gamma _\mu \psi ^b\,,\nonumber\\
  \phi _\mu =& -\ft12 \gamma ^\nu  R'_{\mu \nu }(Q)+\ft{1}{12}\gamma _\mu \gamma
  ^{ab}R'_{ab}(Q)\,,\nonumber\\
  & R'_{\mu \nu }(Q)=\;2D_{[\mu }\psi _{\nu ]}=2\left( \partial _{[\mu }+\ft12b_{[\mu }-\ft32\rmi
A_{[\mu }\gamma _*+\ft14\omega _{[\mu }{}^{ab}\gamma _{ab}\right) \psi
_{\nu ]}\,,\nonumber\\
  f_\mu {}^a =&-\ft14 (R'^{\rm cov})_\mu {}^a +\ft{1}{24}e_\mu {}^a
  (R'^{\rm cov})+\ft14\rmi\tilde R_\mu {}^a (T)\,,\nonumber\\
  & (R'^{\rm cov})_{\mu \nu }{}^{ab}= 2D _{[\mu }\omega _{\nu ]}{}^{ab}-\bar \psi_{[\mu }\gamma ^{ab}\phi _{\nu ]}+\bar \psi _{[\mu } \gamma ^{[a}R_{\nu ]} {}^{b]}(Q)+\ft12\bar  \psi
_{[\mu } \gamma _{\nu ]}R^{ab}(Q)\,,\nonumber\\
& (R'^{\rm cov})= e^\mu _a (R'^{\rm cov})_\mu {}^a\,,\qquad  (R'^{\rm cov})_\mu {}^a= (R'^{\rm cov})_{\mu \nu
 }{}^{ab}e_b^\nu\,,
\label{depfields}
\end{align}
in terms of the independent fields $\{e^{\mu a},\,\psi _\mu ,\,b_\mu,\,  A_\mu\}$, which are the gauge fields of translations, $Q$-supersymmetry, dilatations and $T$-symmetry, the $\U(1)$ $R$-symmetry in the superconformal algebra. In (\ref{depfields}) $D_\mu $ is the covariantization w.r.t. Lorentz transformations, dilatations, and $T$-symmetry, while we use ${\cal D}_\mu $ for the fully covariant derivative w.r.t. all superconformal symmetries.

The field $b_\mu $ can often be omitted since it is the only independent field that transforms under special conformal transformations, which thus implies that it does not appear in the actions.

The Bianchi identity for the curvature $R_{\mu\nu}(P^a)$ is
\begin{equation}
  e_{[\mu }^a R_{\nu \rho ]}(D)= R_{[\mu \nu}^{\rm cov}(M_{\rho ]}{}^a)\,,\qquad R_{\mu \nu }(D) =  R_{a[\mu}^{\rm cov}(M_{\nu]}{}^a)\,.
 \label{BianchiRP}
\end{equation}
Notice that by this relation the third constraint \eqref{eq:constraints} gives the equality
\begin{equation}
R_{\mu \nu }(D)=-\rmi \tilde{R}_{\mu\nu} (T) \label{equalcurv}\,.
\end{equation}
The constraints \eqref{eq:constraints} are not SUSY invariant and therefore modify the algebra. The SUSY transformations of the dependent gauge fields are changed with respect to the transformation that follows from the gauge algebra. The modified variations are
\begin{align}
\delta_{\mathcal{M}}(\epsilon)\omega_{\mu}{}^{ab}&=-\frac{1}{2}\bar{\epsilon}\gamma_{\mu}R^{ab}(Q)\,,\nonumber\\
  \delta_{\mathcal{M}}(\epsilon)\phi_\mu&=-\ft12 \rmi\gamma ^\nu \left(\gamma _* R_{\nu \mu}(T)+\tilde R_{\nu \mu}(T)\right)\epsilon\,,\nonumber\\
\delta_{\mathcal{M}}(\epsilon)f_{\mu}^a&=-\ft18 \bar{\epsilon}\gamma_\mu {\cal D}^b R_b{}^a(Q)\,.  \label{eq:new variations}
\end{align}

\section{Bosonic improved currents}
\label{app:bosonic}

In this section we review the main aspects of CCJ \cite{Callan:1970ze}, and clarify the modifications due to the presence of the $\U(1)$ symmetry. These modifications are automatically generated from the superfield equations. One main ingredient is the chiral scalar curvature superfield ${\cal R}$, which vanish on shell for the conformal case. The bosonic part of this equation are the two complex equations
\begin{equation}
  F^0\approx 0,\, \qquad \ft16R+A_\mu A^\mu +\rmi \nabla ^\mu A_\mu \approx 0\,.
 \label{calR0bosonic}
\end{equation}

\subsection{Review of bosonic conformal currents}

In general for a given Lagrangian ${\cal L}$ one has the energy momentum tensor $T_{\mu\nu}$ by coupling to gravity and varying with respect to $g^{\mu\nu}$. In this way, $T_{\mu\nu}$ is symmetric and conserved.
For example, a neutral scalar field with quartic self-interaction
\begin{equation}
{\cal L}_M=-\ft 12 \partial _\mu \varphi \partial ^\mu \varphi - \lambda \varphi^4\,,
\label{confsimplest}
\end{equation}
has a canonical energy-momentum tensor
\begin{equation}
T_{\mu\nu}=\partial_\mu \varphi \partial_\nu \varphi +g_{\mu\nu}{\cal L}_M\,,
\end{equation}
which is symmetric and conserved, but not traceless.
If the action has dilatational and special conformal symmetry, an improved energy--momentum tensor that is also traceless can be defined \cite{Callan:1970ze}.\footnote{A summary can be found on the webpage of \cite{Freedman:2012zz}, see\\ \href{http://itf.fys.kuleuven.be/supergravity/index.php?id=15&type=ExtraCh15.html}{http://itf.fys.kuleuven.be/supergravity/index.php?id=15\&type=ExtraCh15.html}.}
This is the case for  (\ref{confsimplest}), where
\begin{equation}
\Theta^{\rm c}_{\mu\nu}=T_{\mu\nu}-\frac{1}{6}(\partial_\mu\partial_\nu-g_{\mu\nu}\Box) \varphi^2\,,
\end{equation}
has the property that $\Theta^{\rm c}{}_\mu{}^\mu\approx 0$ on the equations of motions $\Box \varphi\approx 4\lambda \varphi^3$ and has the same charge as $T_{\mu\nu}$.

A conventional gravitation theory is described by an action
\begin{equation}
{\cal S}=\int \rmd^4 x \sqrt{-g}\left[ \ft{1}{2}\kappa^{-2}R +{\cal L}_M\right]\,,
\end{equation}
$\kappa^{-1}=m_{\rm p}$, where ${\cal L}_M$ is the matter Lagrangian, everything except gravity.  If we vary with respect to $g^{\mu\nu}$, we will find
\begin{equation}
R_{\mu\nu}-\frac{1}{2}R g_{\mu\nu}=G_{\mu\nu}\approx\kappa^2 T_{\mu\nu}\,.
\label{Einsteineqn}
\end{equation}
If we want the source of the gravity to be a traceless $\Theta_{\mu\nu}$, and thus to have a conformal and Weyl invariance, we have to replace
$\kappa^{-2}  \to \kappa^{-2}-\tfrac{1}{6}\varphi ^2 $ and consider
\begin{equation}
{\cal S}=\int \rmd^4x \sqrt{-g}\left[\ft12(\kappa^{-2}-\tfrac{1}{6} \varphi^2)R-\ft12 g^{\mu \nu }\partial _\mu \varphi\partial _\nu \varphi -\lambda \varphi ^4 \right]\,.
\label{SphiConfLM}
\end{equation}
The explicit $\kappa $-dependent term obviously is not conformal, but the other terms define a local conformal invariant action.
This leads to a field equation for the graviton:
\begin{equation}
\kappa^{-2}G_{\mu \nu }\approx \Theta_{\mu \nu }^{\rm c}\,,\qquad \Theta_{\mu \nu }^{\rm c}= T_{\mu \nu }-\ft{1}{6}(\nabla _\mu\partial_\nu-g_{\mu\nu}\nabla ^\rho \partial _\rho ) \varphi^2+\tfrac{1}{6} \varphi^2G_{\mu \nu }\,.
 \label{ImprovedGravFE}
\end{equation}
These formulations can be obtained from a conformal action, containing apart from the physical field $\varphi $ also a compensating scalar $\varphi _0$. both have then Weyl weight~1, and one considers the conformal-invariant action with negative kinetic term for the compensator:
\begin{align}
 {\cal S}=&\int \rmd^4 \sqrt{g}\left[-\ft12\varphi _0\Box^C \varphi_0+\ft12\varphi\Box^C\varphi +\lambda \varphi ^4\right]  \nonumber\\
  =&\int \rmd^4 \sqrt{g}\left[\ft12\partial _\mu \varphi_0\partial^\mu  \varphi_0 - \ft12\partial _\mu \varphi\partial^\mu  \varphi+\ft1{12}(\varphi _0^2-\varphi ^2) R  +\lambda \varphi ^4\right]\,.
\end{align}
The Einstein gauge means that we take a gauge choice for dilatation that fixes the constant in front of $R$ to be $(2\kappa ^2)^{-1}$, i.e.
\begin{equation}
  \left.\varphi _0^2\right|_{\poinc}=\varphi ^2 + 6 \kappa ^{-2}\,.
 \label{phiEinstein}
\end{equation}
We obtain the action\footnote{A parametrization that is similar to the main part of the paper is using the variable $s$ with $\varphi =\varphi _0 s$.
In this parametrization the Lagrangian is
\[\frac12\kappa ^{-2}R -\frac12 6\kappa ^{-2}\frac{\partial _\mu s\partial ^\mu  s}{(1-s^2)^2}+ \lambda(6\kappa ^{-2})^2 \frac{s^4}{(1-s^2)^2}\,.\]
}
\begin{equation}
   {\cal S}=\int \rmd^4 \sqrt{g}\left[\frac12\kappa ^{-2}R -\frac12\frac{\partial _\mu \varphi\partial^\mu  \varphi}{1+\ft16\kappa ^2\varphi ^2}+\lambda \varphi ^4\right]
 \label{SphiEinstein}
\end{equation}
The Einstein equation is of the form  (\ref{Einsteineqn}), but $T_{\mu \nu }$ is not traceless, and the action (\ref{SphiEinstein}) does not seem to have a conformal part. The generalization to many real fields is the coset $\SO(1,n)/\SO(n)$, Which is a subcoset of $\SU(1,n)/\SU(n)$, appearing in the related K\"{a}hler couplings to be discussed below.

Conformal gauge means that we put
\begin{equation}
  \left.\varphi _0^2\right|_{\poinc}= 6 \kappa ^{-2}\,.
 \label{phiConf}
\end{equation}
Then the action is of the conformal form (\ref{SphiConfLM}) with traceless energy-momentum tensor as in (\ref{ImprovedGravFE}).

For general scalar couplings, the action has special conformal symmetry if the transformation under dilatations is of the form of a closed homothetic Killing vector \cite{Sezgin:1995th}
\begin{equation}
  \delta \varepsilon ^i = k_{\rm D}^i(\varphi )\,,\qquad \nabla _j k_{\rm D}^i= w\, \delta _j^i\,,
 \label{kD}
\end{equation}
where $w$ is called the Weyl weight, and the covariant derivative uses the connection related to the metric defined by the kinetic terms of the scalars.
The Lagrangian should have weight~4 counting spacetime derivatives as weight~1. Therefore for a sigma model, the weight of the scalars should be~1. Usually we consider scalars that transform as $\delta \varepsilon ^i = \varphi ^i$, and the condition for special conformal symmetry reduces to
\begin{equation}
  \Gamma ^i_{jk}\varphi ^k=0\,.
 \label{Gammaphi0}
\end{equation}

\subsection{Conformal K\"{a}hler couplings: Conformal gauge}

The condition (\ref{Gammaphi0}) is satisfied for K\"{a}hler models with scalars $S^i$ and $S^{\ib}$ if the K\"{a}hler metric $g_{i\jb}$ satisfies
\begin{equation}
  g_{i\jb,k}S ^k=0\,,
 \label{ghomogrank0}
\end{equation}
which is the requirement that the K\"{a}hler potential is homogeneous of degree~1 in $S$ (and the same in $\bar S$), up to a K\"{a}hler transformation.

Such conformal K\"{a}hler models have automatically also a $\U(1)$ Killing vector\footnote{Mathematically defined by the complex structure as $k_{\rm T}^i = J^i{}_j k_{\rm D}^j$.}
\begin{equation}
  \delta S^i = \rmi S^i \lambda _{\rm T}\,,\qquad \delta \bar S^{\ib} = -\rmi \bar S^{\ib} \lambda _{\rm T}\,.
 \label{U1Kahler}
\end{equation}
We consider from \cite{Cremmer:1982en}
\begin{equation}
  {\cal L}= \kappa ^{-2}\sqrt{-g}\left[-\ft16\Phi \, R+3 A_\mu A^\mu  -\Phi _{i\jb}D_\mu  S^i D^\mu  \bar S^{\jb}\right]\,,\qquad D_\mu S^i = (\partial_\mu-\rmi A_\mu ) S^i\,,
 \label{LSbarSPhi}
\end{equation}
where $A_\mu $ is the gauge field of the symmetry (\ref{U1Kahler}), and we take a K\"{a}hler potential $\Phi (S,\bar S)$ that satisfies the above requirements:
\begin{equation}
  3\Delta K \equiv  S^i\Phi _i - \Phi -3 =0\,.
 \label{DeltaKz}
\end{equation}
The field equation for the metric is
\begin{equation}
  -\ft16\Phi G_{\mu \nu }+\ft16\left(\nabla _{(\mu }\partial _\nu -g_{\mu \nu }\Box\right)\Phi +3 A_\mu A_\nu -\ft32g_{\mu \nu }A_\rho A^\rho - \Phi_{i\jb}\left(D_{(\mu }S^iD_{\nu )}\bar S^{\jb}-\ft12D_\rho S^iD^\rho \bar S^i\right)\approx 0\,.
 \label{gmunufieldeqconf}
\end{equation}
Splitting $\Phi $ in $\Phi = -3 +3\Phi _{\rm M}$, where  $\Phi _{\rm M}$ is homogeneous of degree 1 in $S$ and $\bar S$, we can write this as
\begin{align}
 G_{\mu \nu }+6A_\mu A_\nu -3g_{\mu \nu }A^\rho A_\rho \approx \Theta _{\mu\nu}^c=&\Phi_{M i\jb}\left(2  D_{(\mu} S^i D_{\nu)} \bar S^{\jb} - g_{\mu\nu}D^\lambda S^i D_\lambda \bar S^{\jb}  \right)\nonumber\\
 &-\left(\nabla _\mu \partial_\nu -g_{\mu\nu} \Box  \right) \Phi_{M} + G_{\mu\nu} \Phi_{M}
 \label{improvedThetaA}
\end{align}
This improved stress tensor satisfies
\begin{equation}
  \Theta_\lambda ^{c \, \lambda}\approx 0\,.
 \label{Thetacll0}
\end{equation}
Note that the trace of the left-hand side is proportional $R+6A_\mu A^\mu $. It is vanishing corresponds thus to the real part of the second equation in (\ref{calR0bosonic}). The $A_\mu$ field equation
\begin{equation}
  -2\Phi A_\mu + \rmi\left(\Phi_{\ib}\partial ^\mu \bar S^{\ib}-\Phi _i\partial ^\mu S^i\right)\approx 0\,,
 \label{Amubosfe0}
\end{equation}
can be written as
\begin{align}
  8 A_\mu \approx J_\mu\,,\qquad J_\mu=&4\rmi\left(\Phi _{{\rm M}\,i}D ^\mu S^i - \Phi_{{\rm M}\,\ib}D ^\mu \bar S^{\ib} \right)\nonumber\\
  =&  4\rmi\left(\Phi _{{\rm M}\,i}\partial ^\mu S^i - \Phi_{{\rm M}\,\ib}\partial ^\mu \bar S^{\ib} \right)+ 8 A_\mu\Phi_{M}\,.
\label{Amubosfe}
\end{align}
This leads to
\begin{equation}
  \nabla  ^\mu A_\mu\approx \nabla  ^\mu J_\mu \approx 0\,,
 \label{DA0bos}
\end{equation}
which corresponds to the imaginary part of (\ref{calR0bosonic}).

Thus we see that the improved energy-momentum tensor and improved $\U(1)$ currents get a modification w.r.t. the quantities in CCJ, related to the $R$-symmetry. The form of the improved quantities (\ref{improvedThetaA}) and (\ref{Amubosfe}) is very similar to (\ref{ImprovedGravFE}), where the matter current has each time a part proportional to the gravity current.

\subsection{Conformal K\"{a}hler couplings: Einstein gauge}
The bosonic part of the matter-sugra coupled action (without superpotential and $F^I=0$) was given in (\ref{action1}). Using the $S$-variables with the split $X^i= X^0 S^i$, this is
\begin{align}
{\cal S}=\int \rmd^4x\,3\sqrt{-g} &\left[ \ft16 R X^0\bar X^{\bar 0}(1-\Phi _{\rm M})
-X^0\bar X^{\bar 0}\Phi _{{\rm M}\,i\jb} D^\mu S^i D _\mu \bar S^{\jb}
\right.\nonumber\\
&\left.+  (1-\Phi _{\rm M}) D ^\mu X^0D _\mu \bar X^{\bar 0}-\left(X^0\Phi _{{\rm M}i}D_\mu S^i {\cal D}^\mu \bar X^{\bar 0}+\hc\right)
\right].
\label{SbosonicwithA}
\end{align}
The elimination of $A_\mu $ gives (already assumed that $X^0$ will be real)
\begin{equation}
  A_\mu  = \frac{\rmi}{2(1-\Phi _{\rm M})}\left(\Phi _{{\rm M}i}\partial _\mu S^i -\Phi _{{\rm M}\ib}\partial _\mu S^{\ib}\right)\,.
 \label{Amusolution}
\end{equation}
Thus the action after this step becomes
\begin{align}
{\cal S}=\int \rmd^4x\,3\sqrt{-g} &\left[ \ft16 R X^0\bar X^{\bar 0}(1-\Phi _{\rm M})
-X^0\bar X^{\bar 0}\Phi _{{\rm M}\,i\jb} \partial _\mu S^i\partial ^\mu \bar S^{\jb}
\right.\nonumber\\
&\left.  +(1-\Phi _{\rm M}) \partial  _\mu X^0\partial  ^\mu \bar X^0-\left(X^0\Phi _{{\rm M}i}\partial _\mu S^i \partial ^\mu \bar X^0+\hc\right)\right.\nonumber\\
&\left.+\frac{1}{4(1-\Phi _{\rm M})}X^0\bar X^{\bar 0}\left(\Phi _{{\rm M}i}\partial _\mu S^i-\Phi _{{\rm M}\ib}\partial _\mu \bar S^{\ib}\right)^2
\right].
\label{actionA}
\end{align}
In Einstein gauge, we put (and $X^0=\bar X^0$)
\begin{equation}
  X^0\bar X^{\bar 0}(1-\Phi _{\rm M})= \kappa ^{-2}\qquad \rightarrow \qquad \partial _\mu X^0 =\frac{\kappa ^{-1}}{2(1-\Phi _{\rm M})^{3/2}}\left(\Phi _{{\rm M}i}\partial _\mu S^i +\Phi _{{\rm M}\ib}\partial _\mu \bar S^{\ib}\right)\,,
 \label{X0Einstein}
\end{equation}
and this brings the action in the form
\begin{align}
{\cal S}=\int \rmd^4x\,3\sqrt{-g} &\left[ \ft16 R \kappa ^{-2}
-\frac{\kappa ^{-2}}{1-\Phi _{\rm M}}\Phi _{{\rm M}\,i\jb} \partial _\mu S^i\partial ^\mu \bar S^{\jb}
\right.\nonumber\\
&\left.-\frac{\kappa ^{-2}}{4(1-\Phi _{\rm M})^2}\left(\Phi _{{\rm M}i}\partial _\mu S^i+\Phi _{{\rm M}\ib}\partial _\mu S^{\ib}\right)^2 \right. \nonumber\\
&\left.+\frac{\kappa ^{-2}}{4(1-\Phi _{\rm M})^2}\left(\Phi _{{\rm M}i}\partial _\mu S^i-\Phi _{{\rm M}\ib}\partial _\mu S^{\ib}\right)^2\right]
\end{align}
This gives the well-known K\"{a}hler couplings since
\begin{align}
{\cal S}&=\int \rmd^4x\,\sqrt{-g} \kappa ^{-2}\left[\ft12R -3\left(\frac{\Phi _{{\rm M}\,i\jb}}{1-\Phi _{\rm M}}+ \frac{\Phi _{{\rm M}i}\Phi _{{\rm M}\jb}}{(1-\Phi _{\rm M})^2}\right) \partial _\mu S^i\partial ^\mu \bar S^{\jb}\right]\nonumber\\
&=\int \rmd^4x\,\sqrt{-g} \kappa ^{-2}\left[\ft12R -\partial _i\partial _{\jb}\left(- 3\log (1-\Phi _{\rm M})\right)\partial _\mu S^i\partial ^\mu \bar S^{\jb}\right]\,.
 \label{Kahlermodel}
\end{align}
which gives
\begin{align}
{\cal S}=\int \rmd^4x\,\sqrt{-g}\kappa ^{-2} \left[\ft12 R -{\cal K}_{i\jb} \partial _\mu S^i\partial ^\mu \bar S^{\jb}\right]\,,\quad {\cal K}= -3\log (1-\Phi _{\rm M}))\, , \quad {\cal K}_{i\jb}=\partial_i \partial_{\jb}  {\cal K} .
 \label{Kahlermodelfinal}
\end{align}
The matter part is clearly not conformal.

If we include the potential contributions in (\ref{action1}) and eliminate auxiliary fields $F^i$ we get (see $W_i(X)= (X^0)^2 W_i(S)$)
 \begin{align}
{\cal S}=\int \rmd^4x\,\sqrt{-g} \kappa ^{-2} \left[\ft12R - {\cal K}_{i\jb} \partial ^\mu S^i\partial _\mu \bar S^{\jb} -  \frac{\kappa^{-2}}{3(1-\Phi _{\rm M})^2} W_i(S) (\Phi_{{\rm M}i\jb})^{-1} W_{\jb}(\bar S)\right]\,,
 \label{Kahlermodelfinalpot}
\end{align}
and we will create also quartic terms in the Lagrangian.
This is the conformal invariant Wess-Zumino model coupled to supergravity \cite{Wess:1974kz}, which in the rigid limit is the supersymmetric $\varphi^4$ theory. Note that in the conformal gauge we did not eliminate $A_\mu $ and got to a K\"{a}hler model, while in the Einstein gauge we have to eliminate $A_\mu $ to obtain K\"{a}hler kinetic terms.

Since ${\cal K}=-3\log (1-\Phi _{\rm M})$ then $\Phi _{\rm M}=1-e^{-{\cal K}/3}$ and the first condition on $\Phi_{\rm M}$ in \eqref{restrictPhiM} becomes
\begin{eqnarray}
{\cal K}_i S^i ={\cal K}_{\ib }S^{\ib}=3(e^{{\cal K}/3}-1)\, . \label{conditionsonK}
\end{eqnarray}
This gives the condition on the function ${\cal K}$ to be conformal invariant.

\section{Components of superfields from field equations}
\label{ss:methodfe}
We first derive a general result on transformations of field equations in a symbolic form (DeWitt notation). The invariance of the action is the statement
\begin{equation}
  \delta {\cal S}=(\delta \phi ^i) {\cal S}_i=0\,, \qquad {\cal S}_i = \frac{\dl {\cal S}}{\delta \phi ^i }\,,
 \label{invS}
\end{equation}
where $\phi ^i$ are all the independent fields. For the case that we treat in the bulk of the paper, these are\footnote{We have chosen the inverse frame field as basic field such that it is field equation gives directly the Einstein tensor.}
\begin{equation}
  \left\{\phi ^i\right\}=\left\{e^{\mu a},\,\psi _\mu ,\,b_\mu,\,  A_\mu ,\, X^I,\,\Omega ^I,\, F^I,\, \bar X^{\bar I},\,\Omega ^{\bar I},\, \bar F^{\bar I}\right\}\,.
 \label{phiiset}
\end{equation}
This implies for the transformation of the field equation
\begin{equation}
\delta {\cal S} _i  =\delta \phi ^j  {\cal S}_{ji}=\frac{\dl}{\delta \phi ^i} \left(\delta \phi ^j{\cal S}_j\right) -\frac{\dl}{\delta \phi ^i} \left(\delta \phi ^j\right){\cal S}_j = -\frac{\dl}{\delta \phi ^i} \left(\delta \phi ^j\right){\cal S}_j\,.
 \label{transfofe}
\end{equation}
This gives an easier way of deriving the transformation of a field equation in terms of other field equations determined only from the transformation rules without the need of the explicit action.

A second ingredient is covariance. Field equations in general are not yet covariant, but can be `covariantized'. That means that there is for every field $\phi ^i$ a covariant expression $\Theta (\phi )_i$  \cite{Vanhecke:prep} such that the following two sets of equations are equivalent
\begin{equation}
  {\cal S}_i\approx 0\qquad \mbox{and}\qquad \Theta (\phi )_i\approx 0\,.
 \label{approxequiv}
\end{equation}

As a first step in the construction, one has  should consider a coordinate scalar. We indicate this as $T_i$. E.g. for a vector $A^\mu $ one defines
\begin{equation}
  T(A)_a = e^{-1}e_a^\mu  {\cal S}(A)_\mu \,,\qquad {\cal S}(A)_\mu=\frac{\delta {\cal S}}{\delta A^{\mu}} \,.
 \label{ThetaAA}
\end{equation}
This is in general not yet covariant, but it is proven in \cite{Vanhecke:prep} that there exists a covariant expression of the form\footnote{The proof assumes that transformation laws of fields $\delta \phi ^i$ contain at most first order spacetime derivatives.}
\begin{equation}
  \Theta _i = T_i - B _\mu ^A H_{A i}{}^{\mu j}T_j + {\cal O}(B _\mu B _\nu )\,,
 \label{ThetaifromH}
\end{equation}
where the sum over $A$ concerns all standard gauge transformations, $B_\mu ^A$ are the associated gauge fields. The contribution $H$ comes from fields that transform into spacetime derivatives of other fields
\begin{equation}
  \epsilon ^A H_{A i}{}^{\mu j} = \frac{ \pl \delta (\epsilon ) \phi ^j}{\partial (\partial _\mu \phi ^i)}\,.
 \label{defHgeneral}
\end{equation}
For this paper, only $Q$-supersymmetry transformations of fields in (\ref{phiiset}) depend on derivatives of fields, and thus the $A$ refers only to the spinor index $\alpha $ of supersymmetry with $B_\mu ^A\rightarrow\psi _\mu ^\alpha $ (see the remark on the use of the spinor index at the end of this section).

For fermions we use the notation where upper spinor indices $\alpha $ (we make here no difference between chiral and antichiral) refer to the components of the (Majorana) conjugate spinor as explained in \cite[Sec. 3.2.2]{Freedman:2012zz}. This implies that for a spinor $\chi$
\begin{equation}
  \Theta _\alpha = e^{-1}\frac{\dl {\cal S}}{\partial \chi ^\alpha }+\ldots
 \label{Thetaalpha}
\end{equation}
refers to the derivative w.r.t. $\bar \chi $ and this is thus a spinor quantity. On the other hand
\begin{equation}
  \frac{\dr {\cal S}}{\delta \chi_\alpha  }= \frac{\dr {\cal S}}{\delta \chi^\beta }{\cal C}^{\beta \alpha }= - \frac{\dl {\cal S}}{\delta \chi^\beta }{\cal C}^{\beta \alpha }= {\cal C}^{\alpha\beta }\Theta _\beta \,,
 \label{Thetaright}
\end{equation}
is $\overline{\Theta }$, the Majorana conjugate spinor of $\Theta $.

The statement that $\Theta _i$ is covariant means that its transformation contains no spacetime derivatives of the transformation parameter $\epsilon $ and gauge fields appear only hidden in covariant derivatives or covariant curvatures.
There are thus in general in (\ref{ThetaifromH}) correction terms with higher orders of the gauge field (and still proportional to other field equations). However, our goal is to calculate $\delta \Theta _i$, and the transformation of these extra terms will leave always at least one explicit gauge field. The transformation of the higher order terms leads always to terms with at least one explicit gauge field, and thus they will not be relevant to calculate $\delta \Theta _i$.
The first correction term in (\ref{ThetaifromH}) is relevant if the gauge field $B _\mu ^A $ transforms in a non-gauge field.\footnote{Non-gauge field refers here to the `standard gauge transformations', which does not include the frame field $e_\mu ^a$. See \cite[Sec. 11.3]{Freedman:2012zz} for more details.} In our case this is only the case when we calculate the transformation under $S$-supersymmetry, since
\begin{equation}
  \delta (\eta )\psi _\mu  = - \gamma _\mu \eta \,.
 \label{deletapsi}
\end{equation}
For $Q$ supersymmetry in the ${\cal N}=1$ superconformal theory, there are no such terms (there would be for ${\cal N}>1$) and the transformation of $\Theta _i$ can be obtained from (\ref{transfofe}):
\begin{equation}
  \delta (\epsilon )\Theta _i = -\left(\frac{\pl \delta(\epsilon )\phi ^j}{ \partial \phi ^i}\right)_{\rm cov}\,\Theta _j\,,
 \label{delThetacov}
\end{equation}
where `cov' refers to a covariantization of all spacetime derivatives and omission of terms with undifferentiated gauge fields.

The correction term in (\ref{ThetaifromH}) is also important when we consider a covariant derivative on $\Theta ^i$. We then should write
\begin{equation}
  D_a \Theta _i = \left(D_a T_i -\ft12 R_{a\mu }^A H_{A i}{}^{\mu j}  \Theta _j\right)_{\rm cov}\,,
 \label{covDTheta}
\end{equation}
where the derivative to $\partial _\mu \phi ^i$ does not act on the $R_{\mu\nu }^A$. A relevant case that we need below is the calculation of  ${\cal D}^a\Theta (e)_{ba}$, i.e. where the field $\phi ^i$ above is $e^{\nu a}$, see (\ref{Thetadefs}). The correction terms thus originate from transformations of fields that are proportional to $\partial _\mu e^{\nu a}$. These appear only inside the spin connection $\omega _{\mu }{}^{ab}$. In fact, taking into account that we need the derivative w.r.t. the inverse frame field, we have
\begin{equation}
\frac{\partial \omega_\rho{}^{cd}(e)}{\partial \partial_\mu e^{\nu a}}=- \delta_{\rho}^\mu \delta _a^{[c}e^{d]}_\nu + e_{\rho a} e^{\mu [c}e^{d]}_\nu  + g_{\nu \rho } e^{\mu [c}\delta  ^{d]}_a\,.
\label{lowomegadiff}
\end{equation}
Covariant derivatives with spin connection appear in the $Q$-transformations of $\psi _\mu $ and $F_I$. Thus (\ref{covDTheta}) leads to
\begin{equation}
  {\cal D}^a\Theta (e)_{ba}= {\cal D}^aT(e)_{ba}-\ft12 \bar R^a{}_\mu (Q)e_b^\nu  \left(\frac{\partial \bar \delta \bar \psi _\rho}{\partial \partial_\mu e^{\nu a}} \Theta (\psi )^\rho + \frac{\partial \bar \delta F^I}{\partial \partial_\mu e^{\nu a}}\Theta (F)_I\right)\,,
 \label{DaThetaba}
\end{equation}
where $\bar \delta $ refers to a supersymmetry transformation dropping $\bar \epsilon $. Using
\begin{equation}
  \bar \delta \bar \psi _\rho = -\ft14\omega _\rho {}^{cd}\,\gamma _{cd}+\ldots \,,\qquad \bar \delta F^I =\frac{1}{4\sqrt{2}}\omega _\rho {}^{cd}\,\gamma ^\rho \gamma _{cd}+\ldots \,,
 \label{bardelta}
\end{equation}
and (\ref{lowomegadiff}) we obtain
\begin{eqnarray}
   {\cal D}^a\Theta (e)_{ba}= &{\cal D}^aT(e)_{ba}&+\ft18 \bar R_{a\mu} (Q)\left(-\delta _\rho ^\mu \gamma ^a{}_b +e_\rho ^a\gamma ^\mu {}_b +e_{b \rho }\gamma ^{\mu a}\right)\Theta (\psi )^\rho\nonumber\\
   && -\frac1{8\sqrt{2}} \bar R_{a\mu} (Q)
   \left(-\gamma ^\mu   \gamma ^a{}_b +\gamma  ^a\gamma ^\mu {}_b +\gamma _b\gamma ^{\mu a}\right)\Omega ^I\Theta (F)_I
 \label{DaThetaba2}
\end{eqnarray}
Due to the constraints (\ref{eq:constraints}) the last line vanishes, and the first line simplifies to
\begin{equation}
  {\cal D}^a\Theta (e)_{ba}= {\cal D}^aT(e)_{ba}+\ft14 \bar R_{ba} (Q)\Theta (\psi )^a\,.
 \label{DaThetabafinal}
\end{equation}

With the above methods we easily prove that independent of the choice of invariant action
\begin{equation}
  \left\{\Theta (F)_I, \, -\Theta (\Omega )_I,\, \Theta (X)_I\right\}
 \label{Thetachiralmult}
\end{equation}
forms a chiral multiplet of Weyl weight~2. On the other hand, the fields of the Weyl multiplet appear in the transformation laws of the chiral multiplet, such that the multiplet that starts from $\Theta (A)$ involves also the other fields, leading to the result in (\ref{ETM}). To obtain this result, one also needs to use the explicit expressions of the Ward identities (\ref{invS}) for the local symmetries, which are the vanishing of \footnote{$W(K)_a=0$ is the statement mentioned after (\ref{depfields}) that $b_\mu $ does not appear in the action. This is already assumed in the other equations.}
\begin{align}
	\text{Cov. gct: } W(P)_a\equiv&\;{\cal D}^{b}\Theta(e)_{ab}+\Theta(A)^{b}R_{ab}(T)+\ft34\bar{\Theta}(\psi)^{b}R_{ab}(Q)\nonumber\\
	&\;+\left[\Theta(X)^I {\cal D}_a X_I+\bar{\Theta}(\Omega)^I {\cal D}_a \Omega_I+\Theta(F)^I {\cal D}_a F_I+\hc\right]\,,\nonumber\\
	\text{Lorentz: } W(M)_{ba}\equiv&\;\Theta(e)_{[ba]}+\ft14\left[\bar{\Omega}^I\gamma_{ba}\Theta(\Omega)_I+\hc\right]\,,\nonumber\\
	\text{Dilatations: } W(D)\equiv&\;\Theta(e)_a{}^a+\left[X^I\Theta(X)_I+\ft32\Omega^I \Theta (\Omega)_I+2\Theta (F)^I F_I+\hc\right]\,, \nonumber\\
	\text{Spec.conf.: } W(K)_a\equiv&\;\Theta (b)_a\,, \nonumber\\
	\text{$T$-symmetry: } W(T)\equiv&-{\cal D}^a\Theta (A)_a+\rmi\left[X^I\Theta (X)_I -\ft12\overline{\Omega} ^I\Theta (\Omega )_I - 2F^I\Theta (F)_I -\hc\right]\,,\nonumber\\
	\text{$Q$-susy: } W(Q)\equiv&\;{\cal D}^a\Theta(\psi)_a \nonumber\\
	&\;- \ft{1}{\sqrt{2}}\left[\Omega^I\Theta(X)_I+\left(-\slashed{\cal D}X^I+F^I\right)\Theta(\Omega)_I+\slashed{\cal D}\Omega^I \Theta(F)_I+\hc\right]\,,\nonumber\\
  \text{$S$-susy: } W(S)\equiv&\;\gamma _a\Theta (\psi )^a + \sqrt{2}\left[X^I\Theta (\Omega )_I+\hc\right]\,.  \label{WardIdentities}
\end{align}
These are all straightforwardly obtained from (\ref{invS}) by replacing ${\cal S}_i$ with $\Theta _i$ and covariantizing (dropping explicit gauge fields) apart from the one for covariant general coordinate transformations, since this one involves  (\ref{DaThetabafinal}). \footnote{One could expect the same for the $Q$-susy Ward identity, but these extra terms vanish by (\ref{eq:constraints}).}
Using the transformations under covariant general coordinate transformations, which are for the different types of fields
\begin{align}
  \delta _{\rm cgct}e^\mu _a =& -e_b^\mu {\cal D} _a\xi^b \,,\nonumber\\
  \delta _{\rm cgct}\psi _\mu  =& \xi ^a(R_{a\mu }(Q)+\gamma _a\phi _\mu) \,,\qquad \delta _{\rm cgct}A_\mu  = \xi ^aR_{a\mu }(T)\,,\nonumber\\
  \delta _{\rm cgct}X^I =& \xi ^a{\cal D}_a X^I\,,\qquad\delta _{\rm cgct}\Omega ^I = \xi ^a{\cal D}_a \Omega ^I\,,\qquad \delta _{\rm cgct}F^I = \xi ^a{\cal D}_a F^I\,.
 \label{cgct}
\end{align}
Therefore
\begin{align}
  W(P)_a =& {\cal D}^{b}T(e)_{ab}+\Theta(A)^{b}R_{ab}(T)+\bar{\Theta}(\psi)^{b}R_{ab}(Q)\nonumber\\
	&\;+\left[\Theta(X)^I {\cal D}_a X_I+\bar{\Theta}(\Omega)^I {\cal D}_a \Omega_I+\Theta(F)^I {\cal D}_a F_I+\hc\right]\,,
 \label{WPinT}
\end{align}
which due to (\ref{DaThetabafinal}) leads to the expression of $W(P)$ in (\ref{WardIdentities}).

\section{The Poincar\'{e} multiplet}
\label{ss:poincmultiplet}

In  (\ref{ETMGauge}) we presented the Einstein tensor multiplet ${\cal E}_a$ in conformal gauge with the components defined by the conformal transformations (\ref{vectorreal}). The components in \cite{Townsend:1979js,Ferrara:1988qx} have been defined differently, using the Poincar\'{e} transformations defined from the conformal transformations by  (\ref{delPoincare}). A first step to compare the transformations, is to consider the components of the multiplet with Weyl weight~0:
\begin{equation}
  {\cal C}_a^0\equiv {\cal C}_a(X^0\bar X^{\bar 0} )^{-w/2}\,.
 \label{w0multiplet}
\end{equation}
In the case of interest, $w=3$. This is a known procedure, known for multiplets without external indices by \cite{Ferrara:1978jt,Kugo:1982cu}.
The multiplet $(X^0\bar X^{\bar 0} )^{-3/2}$ is in the conformal gauge\footnote{In this appendix, to compare with \cite{Townsend:1979js,Ferrara:1988qx}, we put $\kappa =1$.}
\begin{align}
  C & =1\,,\qquad \zeta =0\,,\qquad {\cal H}= 3 \bar u\,,\nonumber\\
   B_a&= -3 A_a,\,\qquad \lambda =0\,,\qquad  D= \ft92 (u\bar u-A_aA^a)\,.
\label{componentsX0real}
\end{align}
This leads to
\begin{align}
  {\cal C}_a^0  & ={\cal C}_a\,,\qquad {\cal Z}_a^0 ={\cal Z}_a\,,\qquad  {\cal H}_a^0= {\cal H}_a +3 \bar u{\cal C}_a\,,\qquad {\cal B}_{ba}^0= {\cal B}_{ba}-3 A_b{\cal C}_a\,,\nonumber\\
   P_R\Lambda _a^0 & =P_R\Lambda _a+\ft32P_R( \bar u+\rmi\slashed{A}){\cal Z}_a\,,\nonumber\\
    D_a^0&=D_a + \ft92 (u\bar u-A_bA^b){\cal C}_a + \ft32(u{\cal H}_a+ \bar u\,\overline{{\cal H}}_a)+ 3 A^b{\cal B}_{ba}+
   \ft32\rmi \overline{\hat{\phi }} _b\gamma ^b\gamma _*{\cal Z}_a\,.
\label{firststepredef}
\end{align}
These still transform according to (\ref{vectorreal}), now with $w=0$. However, more $S$-supersymmetry terms were absorbed in \cite{Townsend:1979js,Ferrara:1988qx} in redefinitions of the components.
The first case is in the transformation of $P_L{\cal Z}_a^0$. According to (\ref{vectorreal}) with (\ref{delPoincare}) this is for Poincar\'{e} supersymmetry
\begin{align}
  \delta P_{L}\mathcal{Z}_{a}^0=& \ft{1}{2}P_{L}\left(\rmi\mathcal{H}^0_{a}-\gamma^{b}\mathcal{B}^0_{ba}-\rmi\slashed{\mathcal{D}}\mathcal{C}^0_{a}\right)\epsilon+\ft12\rmi P_L \gamma _{ab} {\cal C}^{b0}\left(\rmi\slashed{A}-\bar u\right)\epsilon\,,\nonumber\\
 =& \ft{1}{2}P_{L}\left(\rmi\mathcal{H}^0_{a}-\gamma^{b}\left(\mathcal{B}^0_{ba}+6\eta _{ab}A^cA_c-6 A_bA_a\right)-\rmi\slashed{\mathcal{D}}\mathcal{C}^0_{a}\right)\epsilon-\ft12\rmi P_L \gamma _{ab} {\cal C}^{b0}\bar u\epsilon  \nonumber\\
    & \ft{1}{2}P_{L}\left(\rmi\mathcal{H}^0_{a}-\gamma^{b}\left(\mathcal{B}^{\rm P}_{ba}\right)-\rmi\slashed{\mathcal{D}}\mathcal{C}^0_{a}\right)\epsilon-\ft12\rmi P_L \gamma _{ab} {\cal C}^{b0}\bar u\epsilon
\end{align}
where in the second line we used ${\cal C}_a^0= 6A_a$, and we define then
\begin{equation}
  \mathcal{B}^{\rm P}_{ba}= \mathcal{B}^0_{ba}+6\eta _{ab}A^cA_c-6 A_bA_a= \mathcal{B}_{ba}+6\eta _{ab}A^cA_c-24 A_bA_a\,,
 \label{BabfromB0}
\end{equation}
of which the bosonic symmetric part is the expression (\ref{vanishingB}).

Similarly, $\Lambda _a$ is defined in the transformation of ${\cal H}_a^0$. The conformal covariant derivative of ${\cal Z}_a$, which appears there in (\ref{vectorreal}) contains the $S$-covariantization
\begin{equation}
  {\cal D}_\mu P_L{\cal Z}_a^0 =\ldots  -\rmi \gamma _{ab}{\cal C}^{b0}\phi _\mu \,.
 \label{DmuZphi}
\end{equation}
Furthermore, the $A_\mu $-dependent terms in $\delta {\cal H}_a^0=\ldots +\rmi\bar \eta P_L \gamma _{ab}{\cal Z}^{b0}$ with $\rmi\bar \eta P_L =\ft12\bar \epsilon P_R\slashed{A}+\ldots $ are also absorbed in the definition of the $\Lambda $ in \cite{Townsend:1979js,Ferrara:1988qx}. Therefore we have from these two sources
\begin{equation}
  P_R\Lambda _a^{\rm P}=P_R\Lambda _a^0  -P_R\rmi \gamma ^\mu \gamma _{ab}{\cal C}^{b0}\phi _\mu +\ft12\rmi P_R\slashed{A}\gamma _{ab}{\cal Z}^{b0}\,.
 \label{redefLambda}
\end{equation}
The $D_a$ component appears in the transformation of $P_R\Lambda_a$ with an uncontracted right chiral susy parameter, see (\ref{vectorreal}). In \cite{Townsend:1979js,Ferrara:1988qx} all similar terms are absorbed in the definition of $D_a$. These come from the $S,K$-supersymmetry terms and the transformation of the redefinition terms $P_R(\Lambda _a^{\rm P}-\Lambda _a^0)$. All together they give
\begin{equation}
D_a^{\rm P}=D_a^0+2\varepsilon_{abcd}({\cal D}^b A^c){\cal C}^d+2{\cal B}_{[b}{}^b A_{a]}+\frac32{\cal C}_a A^b A_b+\rmi\bar{\cal Z}^b\gamma_*\gamma_{[a}{\hat{\phi}}_{b]}-\frac14\rmi\bar{\cal Z}^b\gamma_*\gamma_{abc}\hat{\phi}^c\,.
 \label{redefD}
\end{equation}
In summary, the redefinitions that bring components of multiplets following the superconformal transformation laws into the ones following the Poincar\'e transformation laws of \cite{Townsend:1979js,Ferrara:1988qx} are (using ${\cal C}_a=6 A_a$,  ${\cal Z}_a= -6 \hat{\phi}_a$ and the value of the antisymmetric part of ${\cal B}_{ab}$)
\begin{align}
\Delta{\cal H}_a&=18 \bar{u} A_a\,,\nonumber\\
\Delta{\cal B}_{ba}&=-24 A_a A_b+6\eta_{ba}A_{c}A^c\,,\nonumber\\
\Delta P_R\Lambda_a&=\ft{3}{2} P_R(\bar{u}+\rmi\slashed{A}){\cal Z}_a+\frac12\rmi P_R\gamma_c\gamma_{ab}{\cal Z}^b\,A^c+\rmi P_R\gamma^b\gamma_{ac}{\cal Z}_b\, A^c\,,\nonumber\\
\Delta D_a&=\ft{3}{2}(u{\cal H}_a+\bar{u}\,\overline{{\cal H}}_a)+27A_a u\bar{u}+18 A_a A^b A_b+18\varepsilon_{abcd}(\widehat{\cal D}^b A^c) A^d+{\cal B}^{\rm P}_{b}{}^b A_a+2{\cal B}^{\rm P}_{ba}A^b\nonumber\\
&\quad-\ft{1}{4}\rmi\bar{\cal Z}_b\gamma^b\gamma_*{\cal Z}_a+\ft{1}{6}\rmi\bar{\cal Z}^b\gamma_*\gamma_{[b}{\cal Z}_{a]}+\ft{1}{24}\rmi\bar{\cal Z}^b\gamma_*\gamma_{abc}{\cal Z}^c\,,
\label{redefETM}
\end{align}
where e.g. $\Delta{\cal H}_a={\cal H}^{\rm P}_a-{\cal H}_a$, with ${\cal H}^{\rm P}_a={\cal H}^0_a$ and the other P-components are defined in (\ref{BabfromB0}),  (\ref{redefLambda}),  (\ref{redefD}).
Therefore, the result (\ref{ETMGauge}) is in terms of the fields in \cite{Townsend:1979js,Ferrara:1988qx}:
\begin{align}
\mathcal{C}^{\rm P}_{a}=&\;6A_a\,,\nonumber\\
\mathcal{Z}^{\rm P}_{a}=&-6\hat{\phi }_a\,,\nonumber\\
\mathcal{H}^{\rm P}_{a}=&-6\rmi(\widehat{\mathcal{D}}_{a}+3\rmi A_a) \bar u \,,\nonumber\\
\mathcal{B}^{\rm P}_{ba}=&\;3\widehat{G}_{ab}-\eta_{ab}\widehat{G}_c{}^c-6 A_a A_b+3\eta_{ab}A_c A^c+3\eta_{ab}u\bar{u}-3\varepsilon_{bacd}\widehat{\mathcal{D}}^{c}A^d\,,\nonumber\\
P_R\Lambda^{\rm P}_a=&\;2P_R\gamma^b
\left(\widehat{\cal D}_{[a}+\ft{3}{2}\rmi A_{[a}\right){\cal Z}_{b]}+\ft{3}{2} P_R (\bar u + \rmi\slashed{A}){\cal Z}_a+\ft12\rmi P_R\slashed{A}\gamma_{ab}{\cal Z}^b\,,\nonumber\\
\mathcal{D}^{\rm P}_{a}=&-12\widehat{\cal D}_{[b}\widehat{\cal D}^{b} A_{a]}+{\cal B}^{\rm P}_{b}{}^b A_a+2{\cal B}^{\rm P}_{ba}A^b-9u\bar{u} A_a\nonumber\\
&-\frac{1}{6}\rmi\bar{\cal Z}^{b}\gamma_*\gamma_{[b}{\cal Z}_{a]}+\frac{1}{24}\rmi\bar{\cal Z}^b\gamma_*\gamma_{abc}{\cal Z}^c+\rmi\bar{\widehat{R}}_{ab}(Q)\gamma_*{\cal Z}^b\,.
\label{ETMPoncare}
\end{align}
The components in (\ref{ETMPoncare}) agree with the results of \cite{Townsend:1979js,Ferrara:1988qx}.


\providecommand{\href}[2]{#2}\begingroup\raggedright\endgroup

\end{document}